\documentclass[]{aastex631}
\usepackage{amsmath}

\shorttitle{QMC Ensemble of Jupiter Interior Models}
\shortauthors{B. Militzer}
\graphicspath{{./}{figures/}}

\newcommand{\rr}{\ensuremath{\vec{r}}}

\newcommand{\LL}{\ensuremath{\vec{\lambda}}}

\begin{document}

\title{Study of Jupiter's Interior with Quadratic Monte Carlo Simulations}

\author{Burkhard Militzer}
\affil{Department of Earth and Planetary Science, Department of Astronomy,\\ University of
California, Berkeley, CA, 94720, USA}

\begin{abstract}
We construct models for Jupiter's interior that match the gravity
data obtained by the {\em Juno} and {\em Galileo} spacecrafts. To
generate ensembles of models, we introduce a novel {\em quadratic}
Monte Carlo technique that is more efficient in confining fitness
landscapes than affine invariant method that relies on linear
stretch moves. We compare how long it takes the ensembles of walkers
in both methods to travel to the most relevant parameter
region. Once there, we compare the autocorrelation time and error
bars of the two methods. For a ring potential and the 2d
 Rosenbrock function, we find that our quadratic Monte Carlo
technique is significantly more efficient. Furthermore we modified
the {\em walk} moves by adding a scaling factor. We provide the
source code and examples so that this method can be applied
elsewhere. Here we employ our method to generate five-layer models
for Jupiter's interior that include winds and a prominent dilute
core, which allows us to match the planet's even and odd gravity
harmonics. We compare predictions from the different model ensembles
and analyze how much an increase of the temperature at 1 bar and
{\em ad hoc} change to the equation of state affects the inferred
amount of heavy elements in atmosphere and in the planet overall.
\end{abstract}

\section{Introduction} \label{sec:intro}

Since the {\em Juno} spacecraft inserted into orbit around Jupiter in
2016, it has provided us with unprecedented data for the planet's
magnetic field, gravity, and atmospheric
abundances~\citep{Bolton2017}. For this article, the improvement in
the precision of the gravity measurements are particularly
important. While, for example, the gravity harmonic $J_4$ had been
determined to be $J_4 \times 10^6 =-587 \pm 5 $ with data from {\em
  Pioneer} and {\em Voyager} mission, it is now known with much higher
precision,
$J_4 \times 10^6 =‐586.6085 \pm 0.0024 $~\citep{Durante2020}. This has
also led to a revision among the methods and assumptions that go into
modelling the planet's interior
structure~\citep{Stevenson1982,Hubbard2002,HubbardMilitzer2016,Wahl2017a,Ni2018,Nettelmann2021}
but the small error bars have made sampling the available space
  with interior models much more challenging. So here we generate
ensembles of models for Jupiter's interior with a novel Monte Carlo
(MC) method. We employ a number of different model assumption starting
from our reference ensemble of five layer models~\citep{DiluteCore}
that invoke a prominent dilute core that reach out to $\sim$60\% of
the planet's radius as well as contributions from winds that we
derived by solving thermal wind equation~\citep{Kaspi2013} in an
oblate geometry. Interior and wind parameters are optimized
simultaneously, which enabled us to improve upon solutions by
\citet{Wahl2017a} and match the {\em Juno} gravity measurements
exactly.

In our second ensemble we raise the 1 bar temperature to 170~K from
our reference value of 166.1~K that was determined {\em in situ} by
the {\em Galileo} entry probe by matching the temperature-pressure
data points to a dry adiabat~\citep{seiff-1998}. While this fit has a
very small temperature uncertainty, it is not certain to what degree
this measurement represents the planet's global average because the
entry probe fell into a 5 $\mu$m hot spot and thus local weather
effects may have played a role. However, one should not expect
deviations to be too large because radio occultation measurements by
the {\em Voyager} spacecrafts determined the 1 bar temperature to be
165 $\pm$ 5 K~\citep{Lindal1981}. These remote observations very
recently re-analyzed by \citet{Gupta2022} who determined higher
temperatures of 167$\pm$4 and 170$\pm$4~K for latitudes of 6$^\circ$S
and 12$^\circ$N respectively. The temperature increase was primarily
caused by including the chemical species CH$_4$, Ne, Ar, and PH$_3$
when the molecular weight of atmosphere was calculated while the
original value of 165 $\pm$ 5 K was derived by assuming a
hydrogen-helium atmosphere that is free of heavier species.

Given these uncertainties, we constructed an ensemble with
$T_{\rm 1 bar}= 170\,$K here while other authors have consider similar
or even higher values. \citet{Kerley2004} constructed models with
$T_{\rm 1 bar}$= 169~K. Recently, \citet{Nettelmann2021} constructed
models with $T_{\rm 1 bar}$= 175 and 180~K. \citet{Miguel2022} made
the 1 bar temperature a free Monte Carlo parameter and obtained the
best match to the {\em Juno} data while using 1 bar temperatures
between 177 and 188~K. Such a temperature increase may be very
appealing because it increase the entropy of the isentrope and there
lowers the density everywhere in the planet. This makes it easier to
match the {\em Juno} measurements of the gravity coefficients $J_4$
and $J_6$ and more importantly introduces additional flexibilities
into the model to move heavy elements from one layer to another.
Eventually, however, the temperature will be so high that the
isentrope no longer intersect the immiscibility region of
hydrogen-helium mixtures~\citep{Morales2013}, which provides the basis
for the helium rain argument that explains why the {\em Galileo} entry
probe measured Jupiter's atmospheric helium abundance
[$Y/(X+Y) = 0.238 \pm 0.005$~\citet{vonzahn-jgr-98}] to be depleted
compared to the protosolar value of
$Y_0/(X_0+Y_0)=0.2777$~\citep{Lodders2010}. Based on the
semi-analytical equation of state (EOS) by \citet{SC95} and {\em ab
  initio} EOS by \citet{MH13}, we estimate a value for 1 bar
temperature of 180~K for helium rain to have started. However, we
derived this value exclusively with theoretical methods while the
first experimental work, that indirectly inferred the conditions of
H-He phase separation at megabar pressures, placed the onset for this
process at much higher temperatures~\citep{Brygoo2021}.

In our third ensemble, we modify the EOS that we derived with {\em ab
  initio} computer simulations and lowered the density by 3\% \citep{MilitzerHubbard_MOI_2023} in the
pressure interval from 10 to 100 GPa where~\citet{DiluteCore} found
the models to be particularly sensitive. Such {\em ad hoc} EOS
corrections have been introduced many times in the past when the
modeling assumptions by themselves did not yield a good match to the
observations. \citet{Nettelmann2021} lowered the density from 30--200
GPa because without a dilute core nor winds, the {\em Juno} gravity
data could not be reproduced. There is no reason to assume that the
{\em ab initio} EOS calculations are accurate to 1\% level that is
typically assumed to required to model giant planet interiors
accurately. One reason for this level of accuracy is that one aims to
estimate the abundance of heavy elements relative to the protosolar
value of 1.53\%~\citep{Lodders2010}. 

{

To the gravity coefficients $J_4$ and $J_6$, we assume in all three
ensembles that Jupiter's core has been substantially diluted with
hydrogen and helium. The heavily elements, that were essential to
trigger Jupiter's formation, make up only $\sim$18\% by mass. Core
dilution is plausible because {\em ab initio} computer simulations
have shown that all typical core materials such as water, silicates
and iron are soluble in metallic hydrogen at megabar
pressures~\citep{WilsonMilitzer2012,WilsonMilitzer2012b,Wahl2013,Gonzalez2014}. It
is less clear whether the convection in Jupiter's interior is
sufficiently strong to bring up the heavy elements against the forces
of gravity~\citep{Guillot2004}. \citet{Moll2017}, \citet{Muller2020},
and \citet{Helled_2022} studied the interior convection and the
evolution of a primordial, compact core that was originally compose to
100\% of heavy elements. \citet{Liu2019} studied whether Jupiter core
could be diluted by a giant impact. It is conceivable that a small
compact core exists in inside the dilute core but it could not be very
massive because that would take away from the dilute core effect that
enabled us to match $J_4$ and $J_6$. \citet{DiluteCore} placed an
upper limit of 3 Earth masses (1\% of Jupiter's mass) on the compact
core.  
}

Various papers have investigated the effects that different EOSs have
on the inferred properties of
Jupiter~\citep{saumon-apj-04,miguel2016}.  Because there are
uncertainties in the EOS, we constructed ensembles of models for which
we have lower density in a pressure window from $P^*$ to
$10 \times P^*$ and then moved across the entire pressure range of
Jupiter's interior in order to determine on which interval the model
predictions depends most sensitively. It was our goal to provide some
guidance to future experimental and theoretical work on where to
expect the biggest impact for giant planet physics.

We analyze how such EOS change affect the heavy element abundance that
is inferred for the planet's outer envelope. Constructing models with
subsolar or even with a ``negative'' abundance of heavy elements has
enabled previous works to match or nearly match the {\em Juno}
measurements for $J_4$ and $J_6$ without invoking a dilute core or
winds~\citep{HubbardMilitzer2016}. On the other hand if one makes the
assumptions that Jupiter form via core accretion from a well mixed
protosolar nebula, the heavy elements in its atmosphere should occur
in at least solar abundances. The small number of measurements and
remote observations that exist for the atmospheric composition of
giant planets have been reviewed in~\citet{Atreya2019}. With the
exception of neon, the {\em Galileo} entry probe measured the nobel
gases to be three-fold enriched compared to solar. Carbon has found to
to be 4$\times$ solar in Jupiter and 9$\times$ solar in Saturn. If the
same enrichment applied to oxygen and if these measurements were
representative of the Jupiter's entire envelope, it would pose a major
challenge to all modeling activities because most models that match
{\em Juno}'s $J_4$ and $J_6$ only yield heavy elements abundance in
approximately solar proportions. (The same challenge exists for
Saturn, for which typical models~\citep{MilitzerSaturn2019} predict up
to 4$\times$ solar abundance for heavy elements, which is well below
the nine-fold solar measurements for carbon.)

The biggest unknown, however, is the concentration of oxygen, the most
abundant element besides hydrogen and helium. Its abundance informs us
about water which crucial for understanding where and how Jupiter
formed~\citep{HelledLunine2014}. The {\em Galileo} entry probe
measured oxygen to be half solar bringing the total heavy element mass
fraction to 1.7\% before the probe stopped functioning at a pressure
of 22 bar. More recently \citet{Li2020} used {\em Juno's} microwave
measurements to infer an oxygen abundance between one and five times
solar. A more precise determination was not possible because the water
signal is small compared to that of ammonia and its radiative
properties at relevant conditions are not sufficiently well
understood, which provides us with ample motivation to analyze the
amount of heavy element that emerge from our model assumptions.
\\
\\
In this article, we construct three ensembles of model of Jupiter's
interior by introducing a novel Markov chain Monte Carlo methods the
relies on {\em quadratic} rather affine (or linear) moves that are
employed by \citet{Goodman}. We show that our method is more efficient
in confining geometries that are difficult to sample with linear
moves. Since its inception, the affine invariance sampling method has
gained a remarkable level of acceptance in various fields of science
including astronomy and astrophysics where one often needs to
determine posterior distributions of model parameters that are
compatible with observational data that carry uncertainties. For
example, the affine sampling method has been employed to detect
stellar companions in radial velocity catalogues~\citep{PW2018}, to
study the relationship between dust disks and their host
stars~\citep{AR2013}, to examine the first observations of the Gemini
Planet Imager~\citep{MG2014}, to analyze photometry data of Kepler's
K2 phase~\citep{VJ2014}, to study the mass distribution in our Milky
Way galaxy~\citep{MP2017}, to identify satellites of the Magellanic
Clouds~\citep{KP2015}, to analyze gravitational-wave observations of a
binary neutron star merger~\citep{DF2018}, to constrain Hubble
constant with data of the cosmic microwave background~\citep{BV2016},
or to characterize the properties of M-dwarf stars~\citep{MF2015} to
name a few applications. On the other hand,
\citet{HuijserGoodmanBrewer2022} demonstrated that the affine
invariant method exhibits undesirable properties when the multivariate
Rosenbrock density is sampled for more than 50 dimensions.

\citet{Goodman} chose to perform their Markov chain Monte Carlo
simulations with an entire ensemble of walkers (or states) rather than
propagating just a single walker. The distribution of walkers in the
ensemble helps one to propose favorable moves that have an increased
chance of being accepted without the need for a detailed investigation
of the local fitness landscape as the traditional Metropolis-Hastings
Monte Carlo method requires. Many extensions of the
Metropolis-Hastings approach have been
advanced~\citep{AndrieuThoms}. For example, \citet{AdaptiveMH} use the
entire accumulated history along the Monte Carlo chain of states to
adjust the shape of the Gaussian proposal function.

Ensembles of walkers have been employed long before \citet{Goodman} in
various types of Monte Carlo methods that were designed for specific
applications. In the fields of condensed matter physics and quantum
chemistry, ensembles of walkers are employed in {\em variational}
Monte Carlo (VMC) calculations~\citep{martin_reining_ceperley_2016}
that optimize certain wavefunction parameters with the goal of
minimizing the average energy or its variance~\citep{FM99}. Ensembles
are used to vectorize or parallelize the VMC calculations. They are
also employed generate the initial set of configurations for the
walkers in {\em diffusion} Monte Carlo (DMC) simulations. In DMC
calculations, one samples the groundstate wave function by combining
diffusive moves with birth and death processes. An ensemble of walkers
is needed to estimate the average local energy so that the birth and
death rates lead to a stable population size. Walkers with a low
energy are favored and thus more likely to be selected to spawn
additional walkers. Walkers in areas of high energy are likely to died
out.

The birth and death concepts in DMC have a number of features in
common with genetic algorithms that employ a population of individuals
(similar to an ensemble of walkers). The best individuals are selected
and modified with a variety of approaches to generate the next
generation of individuals~\citep{schwefel, Militzer1998}. The
population is needed to establish a fitness scale that enables one to
make informed decisions which individuals should be selected for
procreation. This scale will change over time as the population
migrates towards for favorable regions in the parameter space. This also
occurs in DMC calculations as the walker population migrates towards
regions of low energy, the average energy in the population
stabilizes, and the local energy approaches the ground state energy of
the system.

Ensembles of individuals/walkers are not only employed in genetic
algorithm but are used in many different stochastic optimization
techniques. These methods have primarily been designed for the goal of
finding the best state in a complex fitness landscape, or a state that
is very close to it, rather than sampling a well-defined statistical
distribution function as Monte Carlo method do. Therefore these
optimization are much more flexible than Monte Carlo algorithms that
typically need to satisfy the detailed balance relation for every
move~\citep{Kalos}.

The particle swarm optimization
method~\citep{KennedyEberhart1997,EberhartShi2001} employs an ensemble
(or swarm) of walkers and successively updates their locations
according to a set of velocities. The velocities are updated
stochastically using an inertial term and drift terms
favor migration towards the best individual in the population and/or
towards the global best ever generated.

Furthermore, the downhill simplex method~\citep{numerical_recipes}
employs an ensemble of $N+1$ walkers in $N$ dimensions. The
optimization algorithm successively moves the walker with the highest
or second highest energy in the ensemble in the direction of the
center of mass of the other walkers.  The ensemble of walkers thereby
migrates step by step to more favorable locations in the fitness
landscape without the need to ever compute a derivative of the fitness
function, which makes this algorithm very appealing in situations
where the fitness function is complex and its derivates cannot be
derive with reasonable effort.

In general, efficient Monte Carlo methods are required to have two
properties. They need to migrate efficiently in parameter space
towards the most favorable region. The migration (or convergence) rate
is typically measured in Monte Carlo time (or steps). Once the
favorable region has been reached and average properties among walkers
have stabilized, the Monte Carlo method needs to efficient sample the
relevant parameter space. The efficiency of the algorithm is typically
measured in terms of the autocorrelation time or the size of the error
bars. While in typical applications, algorithms that have fast
migration rates also have a short autocorrelation time, there is no
guarantee that both are linked because the properties of fitness
landscape may differ substantially between the initial and the most
favorable regions of the parameter space. For this reason, we measure
the migration rate and autocorrelation time separately when we
evaluate the performance of the quadratic Monte Carlo method that we
introduce in this article.

This article is organized as follows. In section~\ref{sec:methods}, we
introduce our quadratic Monte Carlo technique and compare it with the
affine invariance method. We also describe how we construct models for
Jupiter's interior.  In section~\ref{sec:results} we present four sets
of results. Frist we compare how the two methods perform for a ring
potential problem and for the Rosenbrock density, then construct
different ensembles of Jupiter's interior, and finally study the
consequences of various corrections to the assumed EOS for the
inferred heavy element abundances in Jupiter's outer molecular
layer. In section~\ref{sec:conclusions}, we conclude. In the appendix,
we show that our quadratic Monte Carlo satisfy the condition of
detailed balance.

\section{Methods} \label{sec:methods}
\subsection{Quadratic Moves}

\begin{figure}[ht!]
\plotone{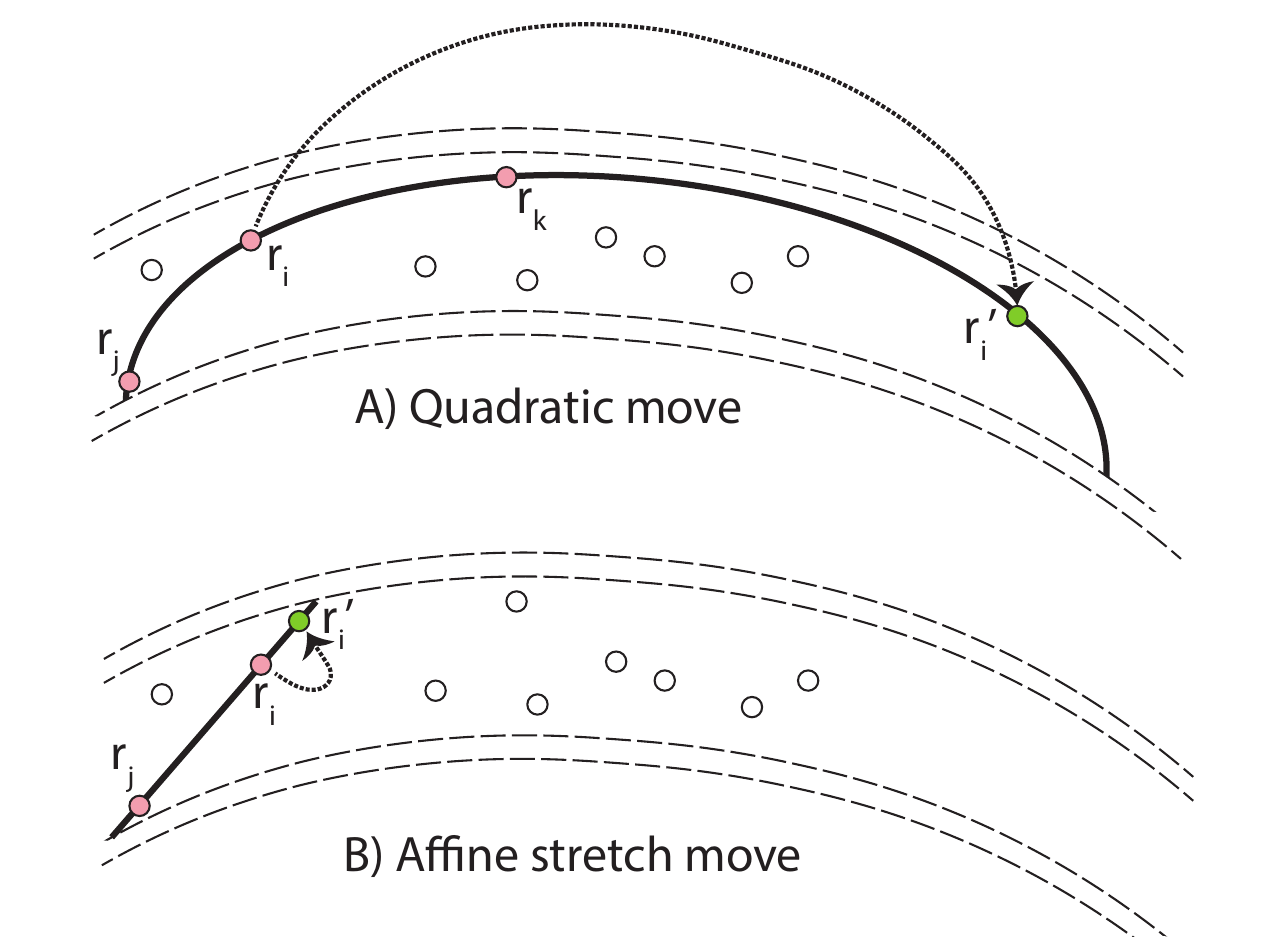}
\caption{Illustration of quadratic and affine stretch moves in a
  confining channel that is represented by the dashed lines. The
  circles indicate the locations of walkers in the ensemble. For the
  quadratic move, two helper points $\rr_j$ and $\rr_k$ are employed
  to sample a new location for walker $i$. Conversely, for the stretch
  move, only one additional point $\rr_j$ is used and walker $i$ may
  thus not travel as far in a single step in a curved channel. \label{fig:moves}}
\end{figure}

We divide our Markov chain MC calculations into $N_b$ blocks, each consisting of
$N_S$ steps. During every step, we attempt to move each of $N_W$
walkers in the ensemble once. A quadratic MC move proceeds as
follows. In addition to the moving walker $i$, we select two other
walkers $j$ and $k$ from the ensemble at random.  Then we perform a
quadratic Lagrange interpolation/extrapolation to sample new
parameters, $\rr_i'$, for walker $i$,
\begin{equation}
\rr_i' = w_i \rr_i + w_j \rr_j + w_k \rr_k
\label{mymove}
\end{equation}
The interpolation weights $w$ are chosen from,
\begin{eqnarray}
w_i &=& L(t'_i \,;\, t_i, t_j, t_k) ,\\
w_j &=& L(t'_i \,;\, t_j, t_k, t_i) ,\\
w_k &=& L(t'_i \,;\, t_k, t_i, t_j) ,\\
L(x \,;\, x_0,x_1,x_2) &\equiv& \frac{x-x_1}{x_0-x_1} \frac{x-x_2}{x_0-x_2} 
\end{eqnarray}
The function $L$ is the typical Lagrange weighting function that
guarantees a proper quadratic interpolation so that $\rr_i'=\rr_i$ if
$t_i'=t_i$; $\rr_i'=\rr_j$ if $t_i'=t_j$; and $\rr_i'=\rr_k$ if
$t_i'=t_k$. We always set $t_j=-1$ and $t_k=+1$ to introduce a scale
into the parameter space, $t$. To satisfy the detailed balance
condition, $T(\rr_i \to \rr_i') = T(\rr_i' \to \rr_i)$, it is key that we sample the parameters
$t_i$ and $t_i'$ from the same distribution $\mathcal{P}(t)$. (We do not set $t_i=0$ but sample it in the same way as $t_i'$.) The acceptance
probability then becomes,
\begin{equation}
  A(\rr_i \to \rr_i') = \min \left[ 1 , \frac{\pi( \rr_i')} {\pi( \rr_i)} \, \left|w_i\right|^N \right]\;.
\label{AQMC}
\end{equation}
The factor $\left|w_i\right|^N$ is needed because we sample the
one-dimensional $t$ space but then switch to the $N$-dimensional
parameter space, $\rr$. It plays the same role as the $\lambda^\alpha$
factor of the affine transformation that we discuss below. In
appendix~\ref{proof}, we derive this factor rigorously from the
generalized detailed balance equation by \citet{GreenMira2001}.

For the sampling distribution, $\mathcal{P}(t)$, one has a bit of
a choice. Our applications have shown that the precise shape is not
important but the width the distribution affects the MC efficiency in
the usual way. If one tries to make too large steps in parameter
space, too many moves are rejected. If the steps are chosen too
small, most moves will be accepted but the resulting states are highly 
correlated and the parameter space is not explored efficiently
either. So we introduce a constant scaling parameter, $a$, that
controls the width of our sampling functions $\mathcal{P}(t)$. Besides
the number of walkers, $N_W$, this is the {\em only} parameter a user
of our quadratic MC method needs to adjust. $a=1.5$ is a perfectly
fine choice. Only if a lot of computer time is to be invested, one
may want to compare the MC efficiency for various $a$ values as we do in
the next section.  For the sampling functions, $\mathcal{P}(t)$, we
propose two
options:\\
a) We sample $t_i$ and $t'_i$, uniformly from the interval
$[-a,+a]$ or\\
b) we draw them independently from a Gaussian distribution that we
center around zero and set the standard deviation equal to $a$.

In Fig.~\ref{fig:moves}, we given an illustration for why quadratic moves
tend to perform well in confined geometries. The move of walker $i$ is
guided by the positions of walkers $j$ and $k$, which both reside in
the narrow channel. Large moves become possible as long as the channel
curvature does not change too rapidly. If it does, one may reduce the
parameter $a$. For $t_i$ values that are sampled from the interval
$[-a,+a]$, the parameter $a$ controls the probability that we choose
the new walker location, $\rr_i'$, by interpolating between $\rr_j$
and $\rr_k$ ($|t_i'|\leq 1$) or by extrapolating from these two
points ($|t_i'| > 1$).

In Fig.~\ref{fig:moves}, we also illustrate the affine stretch moves~\citep{Goodman}
for comparison. To sample the new location for walker $i$, the
position of only one other walker, $j$, is employed to construct this linear transformation,
\begin{equation}
\rr_i' = \rr_j + \lambda (\rr_i - \rr_j)
\label{affine}
\end{equation}
To make such moves reversible, the stretch factor, $\lambda$, must be
sampled from the interval $\left[\frac{1}{a},a\right]$. For the
sampling function, $T(\lambda)$, one has a bit of
choice. \citet{Goodman} followed \citet{Christen} when they chose a function
that satisfies,
\begin{eqnarray}
T_1(\lambda) \;&=&\; \frac{1}{\lambda} \; T_1(\frac{1}{\lambda})\\
T_1(\lambda) \;&\propto&\; \frac{1}{\sqrt{\lambda}} \; \rm{if} \; \lambda \in \left[\frac{1}{a},a\right] \;\;.
\label{T1}
\end{eqnarray}
This function can be sampled by choosing a random number, $\eta$,
uniformly in [0,1] and transforming it according to,
\begin{equation}
\lambda = \frac{(\eta-d)^2}{d^2 a} \;\;\rm{with}\;\; d = \frac{1}{1-a} 
\end{equation}
Alternatively, we can sample $\lambda$ in Eq.~\ref{affine} uniformly
from the interval $\left[\frac{1}{a},a\right]$,
\begin{equation}
T_2(\lambda) = \frac{a}{a^2-1} = \rm{constant~if~} \lambda \,\in\, \left[\frac{1}{a},a\right] {\rm and~T_2(\lambda)=0~elsewhere.}
\label{T2}
\end{equation}
For both sampling functions, a factor,
$\lambda^\alpha= \frac{\left| \rr_i' - \rr_j \right|^\alpha}{\left|
    \rr_i - \rr_j \right|^\alpha}$, must be introduced to the
acceptance probability,
\begin{equation}
  A(\rr_i \to \rr_i') = \min \left[ 1 , \frac{\pi( \rr_i')} {\pi( \rr_i)} \lambda^\alpha  \right]\;.
\label{AAffine}
\end{equation}
For the uniform distribution, $T_2(\lambda)$, one sets $\alpha=N-2$
while one sets $\alpha=N-1$ for $T_1(\lambda)$. Both factors are
caused by the fact that in $N$ dimensions, the area of a sphere around
the anchor point $\rr_j$ scales with
$\left| \rr_i - \rr_j \right|^{N-1}$. The uniform distribution,
$T_2(\lambda)$ already stretches the interval of $a$ values
automatically and therefore $\alpha$ is set to $N-2$ rather than
$N-1$. A derivation for these factors is provided in 
appendix~\ref{proof}.

As a first, very basic test whether any of these methods works
correctly, we applied them to sample the Boltzmann distribution,
\begin{equation}
\pi(\rr) \propto \exp \left\{ - \frac{V(\rr)}{k_B T} \right\}
\label{Boltz}
\end{equation}
for a harmonic potential in $N$ dimensions,
$V(\rr) = \sum_{d=1}^N r_d^2$, in order to verify that the resulting
average potential energy, $\left< V \right>$, agrees with the exact
value of $NT/2$ within error bars. (We set the Boltzmann constant,
$k_B$, to 1 throughout this paper.) This is also a reasonable first
test whether the factors in the acceptance ratios in Eqs.~\ref{AQMC}
and~\ref{AAffine} are set correctly.
As a second test in section~\ref{sec:ring_potential}, we compared the average
potential energy that we obtained with the affine and our quadratic MC
method for a computationally more challenging ring potential.

\subsection{Modified Walk Moves}

\citet{Goodman} also introduced an alternate sampling method: {\it walk}
moves. To move walker $k$ from $\rr_k$ to $\rr_k' = \rr_k + W$, one
chooses at random a subset, $S$, of $N_S$ guiding walkers. $k$ is
excluded from $S$ so that the positions in the subset are independent
of $\rr_k$. The subset size, $N_S$, is a free parameter that one needs to
choose within $2 \le N_S < N_W$. We typically keep $N_S$ constant for
an entire MC chain but we have also performed calculations with a
flexible subset size, for which we selected walkers for the subset
according to a specified probability, $p_S$, but we found no
advantages in using a flexible $N_S$ number over a fixed value.

We follow \citet{Goodman} in computing the average location all walkers in the subset,
\begin{equation}
\left< \rr \right> = \frac{1}{N_S} \sum_{j \in S} \rr_j \quad.
\end{equation}
but we then modify their formula for computing the step size, $W$, by introduding a scaling factor $a$:
\begin{equation}
W = a \sum_{j \in S} Z_j \left( \rr_j - \left< \rr \right> \right) \quad.
\label{eq:W}
\end{equation}
$Z_j$ are univariate standard normal random numbers. By setting $a=1$, one obtains the
original walk moves, for which the covariance of the step size, $W$, is
the same as the covariance of subset $S$. However, the new scaling
parameter, $a$, enables us to make smaller (or larger) steps in
situations where the covariance of the instantaneous walker distribution
is a not an optimal representation of local structure of the sampling
function. We will show later that the scaling factor $a$ enables us to
significantly improve the sampling efficiency of the Rosenbrock
function and for the ring potential in high dimensions.

\subsection{Equation of State} \label{sec:EOS}

The EOS of hydrogen-helium mixtures plays a crucial role in the
modeling Jupiter's interior structure because both gases make up the
bulk of the planet. We derive he EOS by combining the \citet{SC95}
predictions at low pressure and with results from {\em ab initio}
computer simulations at high pressure (P $\ge$ 5
GPa)~\citep{MH13}. For a given composition and entropy, both EOSs
provide a $\rho(P)$ relationship. One can gradually switch from one to
the other as function of pressure. Still there are two primary sources
of uncertainty to consider:

(1) First, current {\em ab initio} calculations are based on the
density functional theory and employed on the PBE
functional~\citep{PBE} while other choices are possible. Currently we
lack experimental data to determine how accurately any of the existing
functionals~\citep{Clay2016} characterize liquid hydrogen at megabar
pressures. X-ray diffraction experiments of solid materials at room
temperature have shown that the PBE functional underestimates the
density of materials by a few \% while the earlier local density
approximation, that was constructed from results by~\citet{CA80},
overestimates the density of solids. However, simulations based on the
PBE functional are in very good agreement with the shock wave
measurements (see \citet{Knudson17} and \citet{MilitzerJGR2016}) that measured the
density of deuterium at megabar pressure more accurately than
previously possible. Still accurate density measurements of liquids
remain a challenged because X-ray diffraction measurements cannot be
applied. On the other hand, quantum Monte Carlo
calculation~\citep{Mazzola2018} have predicted hydrogen to be more
dense than the PBE predictions. However, a higher-than-PBE density
relationship would make the modeling of Jupiter's interior more
difficult and likely lead to subsolar heavy element abundance in the
other envelope, as we will discuss in the results section of this
manuscript.

(2) The second EOS uncertainty arise from the temperature profile of
the isentropes. Locally this is characterized by the Gr\"uneisen
parameter,
$\gamma = - \left. \frac{\partial \ln T}{\partial \ln V}\right|_S = T
\left. \frac{\partial P}{\partial T}\right|_V / \left. \frac{\partial
    E}{\partial T}\right|_V $~\citep{MH08}. Globally, one can state
that the temperature at 1 bar defines an entropy value that determines
$P$-$T$ relationship for the entire thickness of layer in the planet
as long as it is homogeneous and convective. Different predictions
from local and global approaches have led to very different
predictions how hot Jupiter's interior is~\citep{MHVTB,NHKFRB}. With
the global approach, one determine the absolute entropy for a grid of
$\rho$-$T$ point with the thermodynamic integration
method~\citep{Morales2009,Militzer2013} and then find the isentrope
through interpolation. We favor this approach~\citep{MH08b} because
every $S(\rho,T)$ point is independent. So if one particular
calculation were inaccurate, it would not affect the results
elsewhere. With the local approach, one needs a very dense grid EOS
points to numerically compute the derivatives that are needed trace an
isentrope by computing $\gamma$ at every step. A second and important
reason for the disagreement on Jupiter's interior temperature profile
was that the local approach requires a reliably starting point for the
isentrope and {\em ab initio} simulations do not work at 1 bar because
the density is too low.

\subsection{Modeling Jupiter's Interior} \label{sec:interior}

\begin{figure}
\plotone{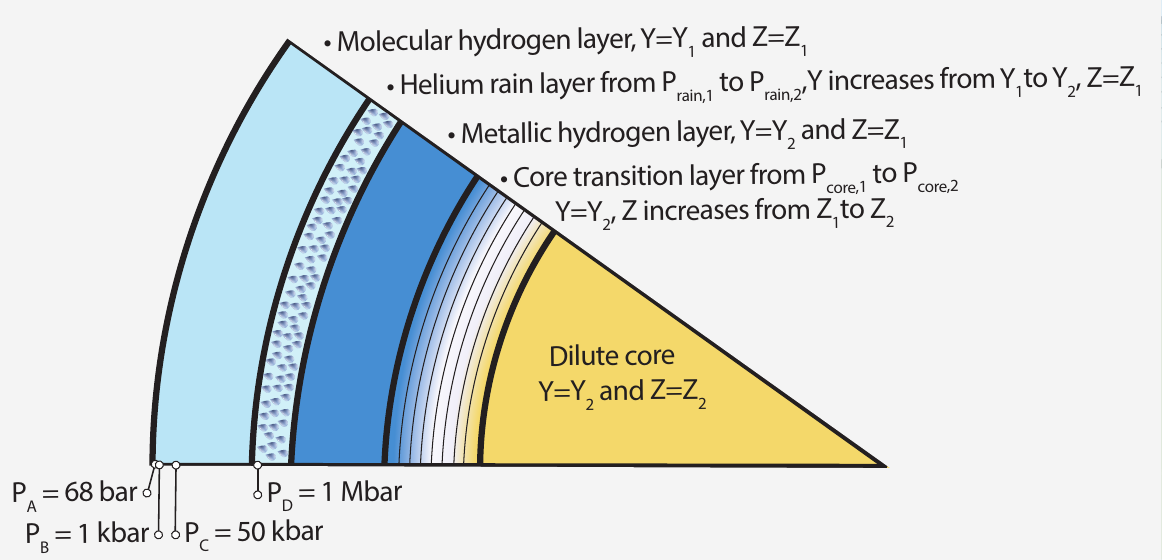}
\plotone{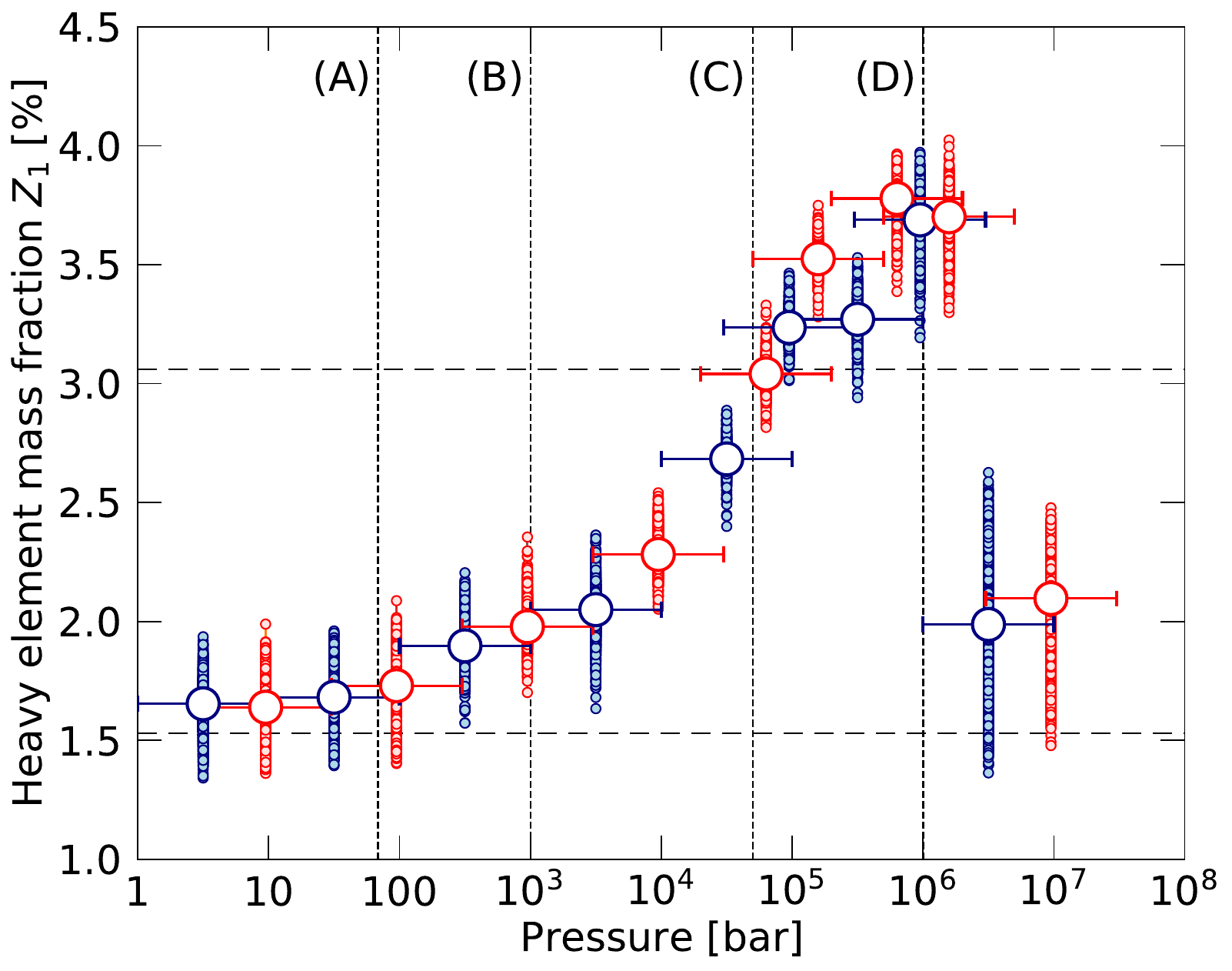}
\caption{Upper panel: Five layer models of Jupiter's interior. Lower panel: Effect of an EOS perturbation on heavy element abundance, $Z_1$. For 18 separate MC calculations, we lowered the density of our H-He EOS by 3\% over a pressure interval from $P^*$ to $10 \times P^*$ and studied how $Z_1$ increased. The small circles show individual $Z_1$ points while the large circle represent the ensemble average. The horizontal bar indicate the interval from $P^*$ to $10 \times P^*$. The horizontal lines mark protosolar and twice protosolar abundances. The vertical lines A trough D are mark pressures of 68, 1005, $5 \times 10^4$, and $10^6$ bar (1 Mbar = 100 GPa) that are also pointed out in the upper panel. (All points were calculate in the same way. Different colors were just introduced for clarity.) 
\label{EOS}}
\end{figure}

We model Jupiter's interior with five distinct layers that we
illustrate in Fig.~\ref{EOS}. The outer layer contains a mixture of
molecular hydrogen, helium, and heavier elements. We derive its EOS by
following \citet{HubbardMilitzer2016}.  We keep the entropy of this layer fixed by
specifying the 1 bar temperature, 166.1 or 170 K. The helium mass
fraction is held constant at the {\em Galileo} value of
$\tilde{Y}_1= Y/(X+Y) = 0.238$~\citep{vonzahn-jgr-98}. $Z_1$
represents the mass fraction of the heavy elements. The parameters
$P_{\rm rain,1}$ and $P_{\rm rain,2}$ mark the beginning and ending
pressures of the helium rain layer where the helium fraction,
$Y/(X+Y)$, gradually rises from $\tilde{Y}_1$ to a higher value
$\tilde{Y}_2$. Following~\citet{DiluteCore}, we adopt this functional
form,
\begin{eqnarray}
\tilde{Y}(P) &=& \tilde{Y}_1 + x^\alpha \left[\tilde{Y}_2-\tilde{Y}_1 \right] \;\;{\rm with}\;\; x = \frac{\log(P/P_{\rm rain,1})}{\log(P_{\rm rain,2}/P_{\rm rain,1})}
\label{rain}
\end{eqnarray}
The $\tilde{Y}_2$ value is adjusted so that the planet overall
(excluding heavy elements) has a helium fraction equal to the
protosolar value of $Y_0/(X_0+Y_0)=0.2777$~\citep{Lodders2010}. Inside
of this layer is a thick, homogeneous, and convective layer of mostly
metallic hydrogen that extends down to the core transition layer. The
parameters $P_{\rm core,1}$ and $P_{\rm core,2}$ determine the
beginning and ending pressures of this layer. We assume it to be
stably stratified because the heavy element fraction increases
gradually from $Z_1$ to $Z_2$. $Z_2$ is the heavy element abundance in
the dilute core, which we assume to be homogeneous and
convective. Together with the metallic hydrogen layer is contribution
to generating Jupiter's magnetic field (see analysis by \citet{Moore_2022}).

To compare the different models, we define the core mass,
$M_{\rm core}$, to be the mass inside of the pressure level,
$P_{\rm core,2}$. The mass of the envelope, $M_{\rm env}$, is the mass
outside the pressure level, $P_{\rm core,1}$. The remaining mass in
between both pressures, is the mass of the core transition layer,
$M_{\rm trans}$.

We employ the concentric Maclaurin spheroid (CMS)
method~\citep{Hubbard2013} to construct a hydrostatic solution of a
uniformly rotating oblate planet and then use the thermal wind
equation to compute the contributions from the zonal winds. The CMS
technique treats the effects of rotation nonperturbatively and is thus
significantly more accurate than the traditional theory of
figures~\citep{ZT1978} that starting from a nonrotating planets and
then adds rotational effects using an expansion of different
orders~\citep{saumon-apj-04,Nettelmann2021}.

We employ our quadratic Monte Carlo method to construct ensembles of
Jupiter models by accepting and rejecting moves according to the
$\exp(-\chi^2/2)$ function that includes four different terms, 
$\chi^2 = \chi^2_J + \chi^2_{\rm H-He} + \chi^2_{\rm wind} + \chi^2_{\rm guide}$. 
The most important one measures the deviations of even and odd gravity
harmonics between model predictions and the {\em Juno}
measurements~\citep{Durante2020},
\begin{equation}
  \chi^2_J = \sum\limits_{i=1}^{10} \left[ \frac{ J_{i}^{\rm model}  - J_{i}^{\rm Juno} }{ \delta J_{i}^{\rm Juno} } \right]^2
  \quad,
  \label{chi_J}
\end{equation}  
where $ \delta J_{i}^{\rm Juno}$ are the 1-$\sigma$ uncertainties
of the measurements.

While Eq.~\ref{chi_J} is certainly the most important model generation
criterion, there are a number of other well motivated constraints to
consider~\citep{MilitzerSaturn2019}. For example, one would want to
favor models with $P_{\rm rain,1}$ and $P_{\rm rain,2}$ value that are
broadly compatible with phase diagram of H-He mixtures as derived by
\citet{Morales2013}. From the assumed molecular and metallic adiabats,
we can infer the temperatures $T_1$ and $T_2$ that correspond to both
pressures. For both pairs $P_{\rm rain,1}$-$T_1$ and
$P_{\rm rain,2}$-$T_2$, we find the closest points on the
immiscibility curve, $P^*_1$-$T^*_1$ and $P^*_2$-$T^*_2$, that
minimize the following immiscibility penalty function,
\begin{equation}
  \chi^2_{\rm H-He} = \sum\limits_{i=1}^{2}C_P \left| \frac{P^*_i-P_i}{P_i}  \right| + C_T \left| \frac{T^*_i-T_i}{T_i}  \right| 
  \quad,
  \label{chi_H-He}
\end{equation}  
before we add the resulting minimum value to the total $\chi^2$. $C_P$
and $C_T$ are weights that must be balanced with those in other
$\chi^2$ terms. We set $C_T/C_P=2$. Implicitly the $\chi^2_{\rm H-He}$
term also introduces a penalty for metallic adiabats that are too hot
to be compatible with the assumed immiscibility curve. We chose not to
square the individual terms in Eq.~\ref{chi_H-He} because there is
currently no agreement between theoretical and experimental results
where in pressure-temperature space, hydrogen and helium become
immiscible. \citet{Vo07} had shown with {\em ab initio} simulation
that hydrogen and helium are miscible at 8000~K. With more careful
{\em ab initio} Gibbs free energy calculations, \citet{Morales2013}
predicted hydrogen and helium to phase separate at approximately
6500~K for a pressure of 1.5 Mbar. Recent shock wave experiments by
\citet{Brygoo2021} that combined Doppler interferometry and
reflectivity measurements placed the onset of immiscibility at a much
higher temperature of 10$\,$200~K at 1.5 Mbar. Based on \citet{MH13},
this corresponds to an entropy of 8.3~k$_B$/electron and imply that
helium rain would set in as soon as a giant planet's 1~bar temperature
cools to 360~K. ({\em Ab initio} methods predict 180~K.) Helium rain
would begin much earlier and cover a longer fraction of a giant
planet's lifetime. \citet{FH04} for example estimated that Jupiter's
1~bar temperature only cooled by 10~K during the last 1.5 billion
years. Also according to \citet{Wahl2021}, helium rain would have
already started on hot exoplanets in 9 day orbits like Kepler-85b but
not on exoplanets in 1 and 3 days orbits such as WASP-12b and
CoRoT-3b. Because the deviations of the {\em ab initio} predictions
are unexpectedly large and these findings to not yet been reproduced
with other laboratory measurements, we will employ the
\citet{Morales2013} results when we evaluate the $\chi^2_{\rm H-He}$
term in Eq.~\ref{chi_H-He} for this manuscript. 
Conversely, \citet{Miguel2022} does not invoke a term like
Eq.~\ref{chi_H-He} or a gradual change as in Eq.~\ref{rain}. Instead
they employ a sharp transition from the molecular to the metallic
hydrogen layer without incorporating predictions from {\em ab initio}
simulations. This transition occur between 2 and 5 Mbar in most
models.

Third we add a penalty term~\citep{DiluteCore},
\begin{equation}
  \chi^2_{\rm wind} = \frac{1}{m} \sum\limits_{i=1}^{m} 
\left\{
\begin{matrix} 
\left[ H(\mu_i)-H_{\rm max}  \right]^2 & {\rm if\,\,} H(\mu_i)>H_{\rm max} \\
0 & \;\;\;\; \;\;\;\; \;\;\;\;\;  {\rm if\,\,} H_{\rm min} \le H(\mu_i) \le H_{\rm max} \\
\left[ H_{\rm min} - H(\mu_i)\right]^2 & {\rm if\,\,} H(\mu_i)<H_{\rm min} \\
\end{matrix}
\right.
\quad,
\label{chi_wind}
\end{equation}  
that keeps the depth of our winds, $H$, within perscribed limits of
$H_{\rm min}=1500\,$km and $H_{\rm max}=4500\,$km to keep them broadly
compatible with earlier predictions~\citep{Guillot2018}. We evaluate
them at $m=61$ equally spaced $\mu$ points between --1 and +1 with
$\mu = \cos(\theta)$ and $\theta$ being the colatitude. We directly
use the observed cloud-level winds from \citet{Tollefson2017} but then
assume the wind depth to be latitude dependent. Alternatively one can
allow the winds on the visible surface to deviate from the
observations and keep the wind depth the same for all latitudes. Both
types of wind solutions are compared in \citet{DiluteCore}. 

We solve the thermal wind equation~\citep{Kaspi2016} to derive the density perturbation, $\rho'$,
\begin{equation}
\frac{\partial \rho'}{\partial s} = \frac{2 \omega}{g}\frac{\partial}{\partial z}\left[\rho u\right]
\quad,
\end{equation}
for a rotating, oblate planet~\citep{Cao2017} in geostrophic
balance. $z$ is the vertical coordinate that is parallel to the axis
of rotation. $s$ is the distance from the equatorial plane along a
path on an equipotential. $\rho$ is static background density and $g$
is the local acceleration. We obtain both from our CMS calculations of
a particular model, which means our wind model and the interior
structure are selfconsistent. $u$ is the differential flow velocity
with respect to the uniform rotation rate, $\omega$.  We represent $u$
as a product of the surface winds, $u_s$, from \citet{Tollefson2017}
and a decay function of $\sin^2(x)$ form from
\citet{MilitzerSaturn2019} that keeps the wind speeds initially
constant before they decay over a small depth interval. {This is
  consistent with assumptions made by \citet{Dietrich2021} and
  \citet{Galanti_Kaspi_2021} while in \citet{Kaspi2018} and
  \citet{Miguel2022} a gradual decay of the wind speed with depth is assumed.  }

We integrate the density perturbation, $\rho'$, to determine the
dynamic contributions to the gravity harmonics before combining them
with the static gravity harmonics that we have obtained from the CMS
calculation, $J_n^{\rm model} = J_n^{\rm static} + J_n^{\rm dynamic}$.
The resulting harmonics are then compared with the {\em Juno}
measurements in Eq.~\ref{chi_J}. We work with the error bars of the
{\em Juno} measurements, $\delta J_{i}^{\rm Juno}$ directly since we
construct selfconsistent models in which wind terms can compensate for
variations in the interior structure. {This is one of the main
  differences to the recent work by \citet{Miguel2022} who performed
  interior and wind calculations separately and increased the {\em
    Juno} error bars by a factor of 30 to represent an unknown
  contribution to even harmonics that comes from the winds.} The other
main difference is that we used the nonperturbative CMS approach while
\citet{Miguel2022} relied on the 4th order theory of figures method
but then compute a correction for a subset of models.

Finally we add the penalty term,
\begin{equation}
  \chi^2_{\rm guide} = C
\left\{
\begin{matrix} 
\left[ p_{\rm min} - p  \right]^2 & {\rm if\,\,} p < p_{\rm min} \\
0 & {\rm otherwise}\\
\end{matrix}
\right.
\quad,
\label{chi_para}
\end{equation}  
that help us guide the Monte Carlo ensemble to reach and remain in
parameter region with $p \ge p_{\rm min}$ that we consider
physical. (Similar terms can assure $p \le p_{\rm max}$.) We find such
a soft approach to work better then a hard constraint that would
reject any model that violates the condition, $p \ge p_{\rm
  min}$. Still we set $C$ to a high value like 1000 to assure
compliance. We verify that $\chi^2_{\rm guide}=0$ for models that we
publish. $Z_1 \ge Z_{\rm protosolar}$ is an obvious condition to
satisfy but we also require $Z_2 \ge Z_1$ and $S_2 \ge S_1$. 

The {\em Juno} gravity
measurements~\citep{Folkner2017a,Iess2018,Durante2020} have reached a
very high degree of accuracy and the fact that we employed the error
bars directly, rather than inflating them, underlines the need to an
efficient sampling method that we provide with our quadratic Monte
Carlo approach.

Some interior parameters are allowed to vary freely during the Monte
Carlo calculations while others are others constrained by
observations. For example, we do not vary the helium fractions, $Y_1$
and $Y_2$ because $Y_1$ is constrained by measurements of the {\em
  Galileo} entropy probe and $Y_2$ is derive so that the planet
overall has a protosolar $\tilde{Y}$. The heavy elements fractions,
$Z_1$ and $Z_2$ are employed so that the model matches the planet's
mass and $J_2$. During the Monte Carlo procedure, we only vary the
four pressures, $P_{\rm rain,1}$, $P_{\rm rain,2}$, $P_{\rm core,1}$,
$P_{\rm core,2}$, the helium rain exponent $\alpha$, the entropy of
the deep interior, $S_2$, and the depth of the winds, $H(\mu_i)$. We
do not introduce a prior distribution or apply any hard constraints to
these four pressure values. Their posterior distribution is just a
result of the different $\chi^2$ terms that we have described in this
section.

\section{Results} \label{sec:results}

\subsection{Application to Ring Potential} 
\label{sec:ring_potential}

\begin{figure}[ht!]
\plotone{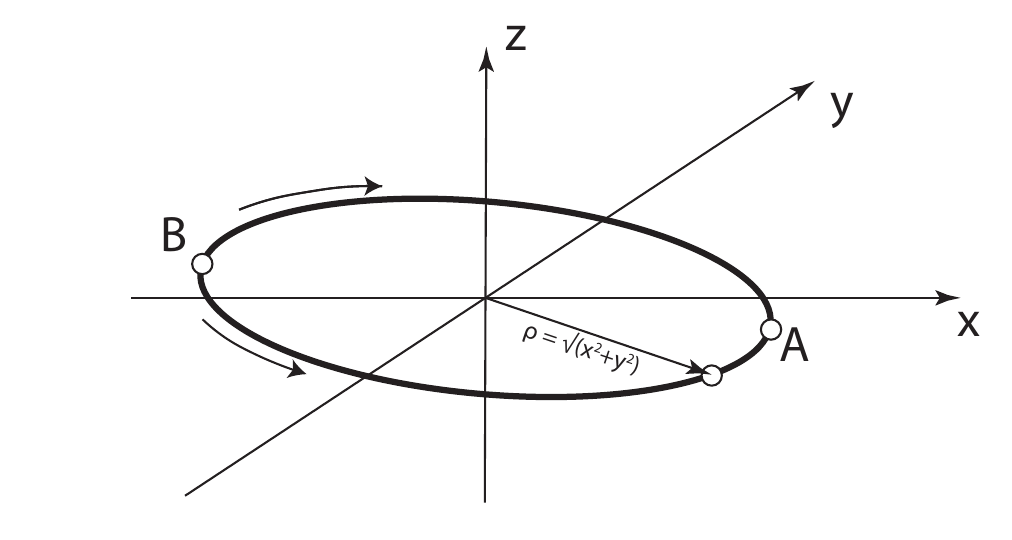}
\caption{Illustration of the ring potential, $V(\rr=(x,y,z))$, that we constructed to study how well different MC algorithms work in confined geometries. By construction, the potential becomes small if the distance $\rho$ equals a given radius, $R$. We slightly tilted the ring to illustrate the effect of the last term in Eq.~\ref{eq:pot} that breaks the axial symmetry by lowering the potential for positive $x$ values. \label{fig:pot}}
\end{figure}

In order to study how our QMC method performs in confined geometries,
we constructed the following ring potential,
\begin{equation}
V(\rr) = (2m)^{2m} \left[ (\rho-R)^{2m} + \sum_{i=3}^{N} r_i^{2m} \right] - C r_1\;\;,
\label{eq:pot}
\end{equation}
where $\rr=\{r_1, \ldots, r_N\}$ is a vector in the $N \ge 2$
dimensional parameter space. $\rho = \sqrt{r_1^2 + r_2^2}$ is the
distance from the origin in the $r_1$-$r_2$ plane. The first term
ensures that the potential is only small along a ring of radius, $R$,
in $r_1$-$r_2$ space as we illustrate in Fig.~\ref{fig:pot}. The
second term keeps the magnitude of all remaining parameters,
$r_{3 \ldots N}$ small. Increasing the positive integer, $m$, allows
us to make the potential more confining by making the potential walls
around the ring steeper. Finally we introduce the last term to break
axial symmetry. Typically we set C to small value like 0.01 so that
the potential minimum is approximately located at point
$\vec{A}=(+R,0, \ldots)$ while the potential is raised at opposing
point $\vec{B}=(-R,0, \ldots)$. The prefactor of the first term in
Eq.~\ref{eq:pot} is introduced so that the location of potential
minimum does not shift much with increasing $m$.

For this test case, we insert the ring potential into the Boltzmann
distribution in Eq.~\ref{Boltz}. If we initialize an ensemble of
walkers in the vicinity of point $\vec{B}$, the algorithm has no
choice but to travel along the ring until it reaches the area of point
$\vec{A}$ where the sampling probability is highest, the most relevant
states will be sampled, and only then the block averages will start to
stabilize.

\begin{figure}[ht!]
\plotone{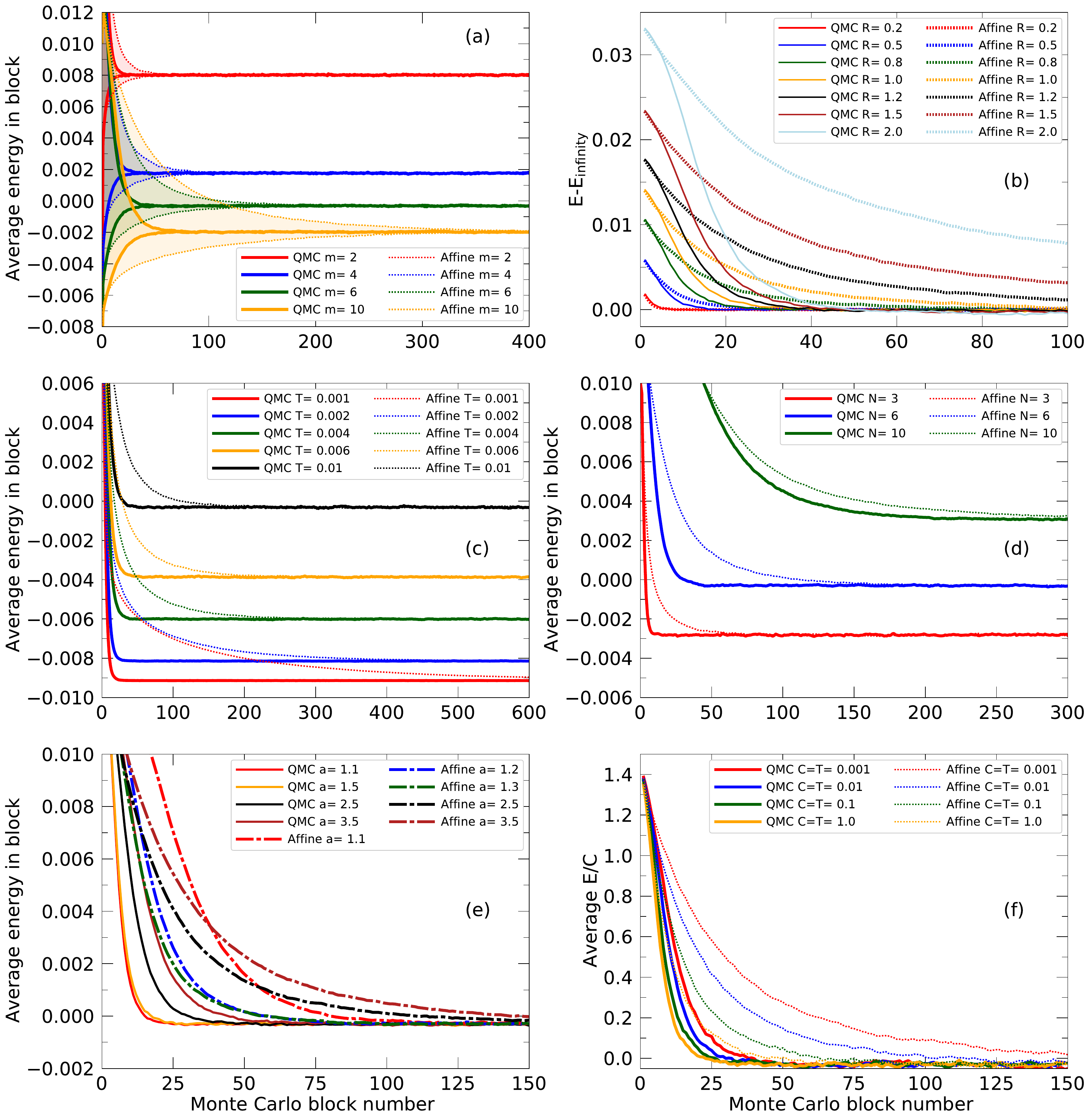}
\caption{Performance comparison between affine MC and our quadratic MC
  methods. The average potential energy in a MC block is plotted as
  function of block number in order to illustrate how long it takes for
  either method to converge. For all parameters considered here, the
  QMC method does so more efficiently. In panel (a), two curves are 
plotting for every method. Tose that converge from above represent 
MC ensembles that were initialized from the high-energy point $\vec{B}$ 
(see Fig.~\ref{fig:pot}) while those converge from below where started 
from the low-energy point $\vec{A}$. In all following panels, we only 
show ensembles that were initialized near $\vec{A}$.
  In panel (b), the final, converged energy has been
  subtracted for clarity. In panel (f), the energy has been divided by
  constant $C$. (To reduce the noise, 1000 independent MC simulations
  have been averaged to generate each curve.) \label{fig:ringResults}}
\end{figure}

\begin{figure}[ht!]
\gridline{\fig{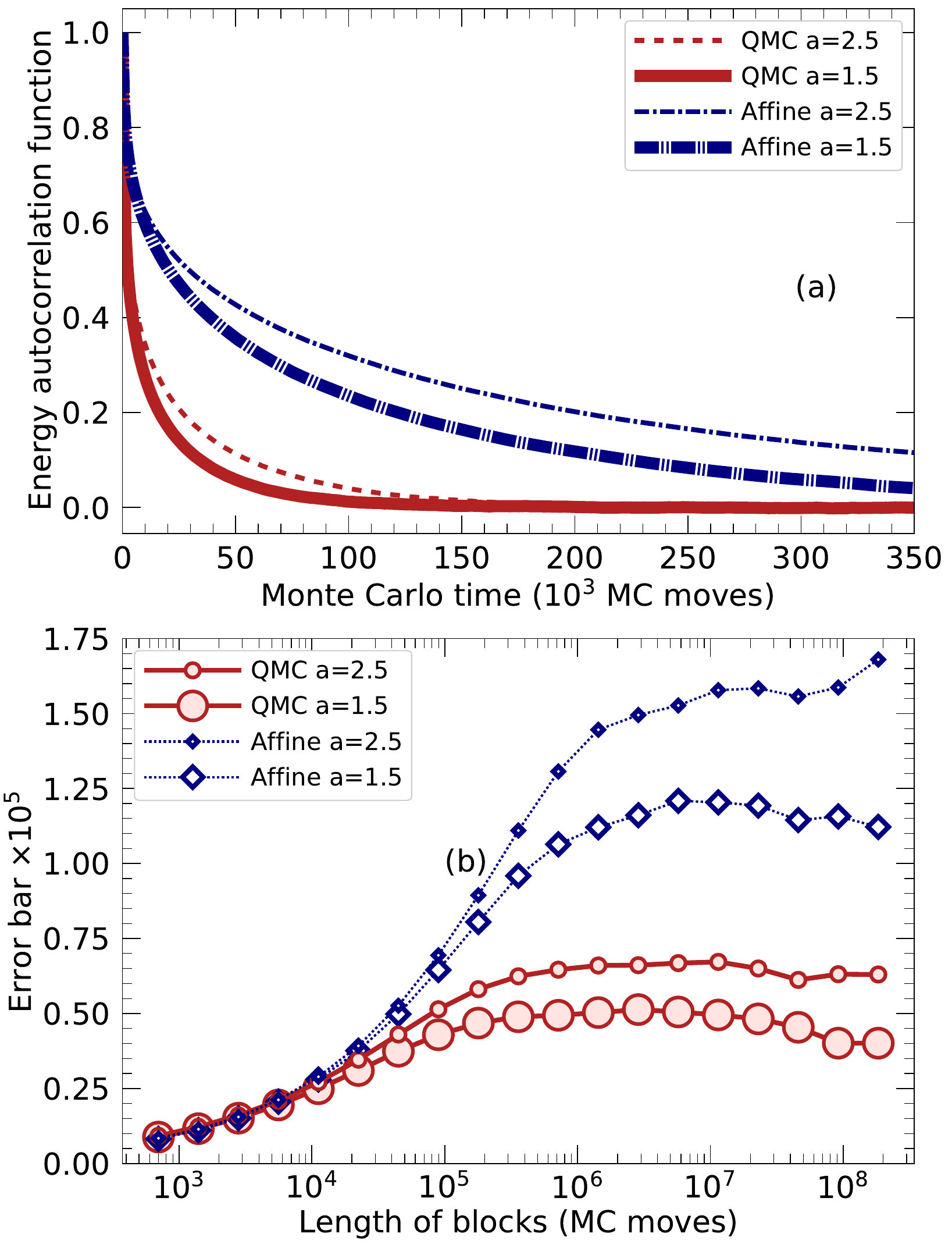}{0.5\textwidth}{}
          \fig{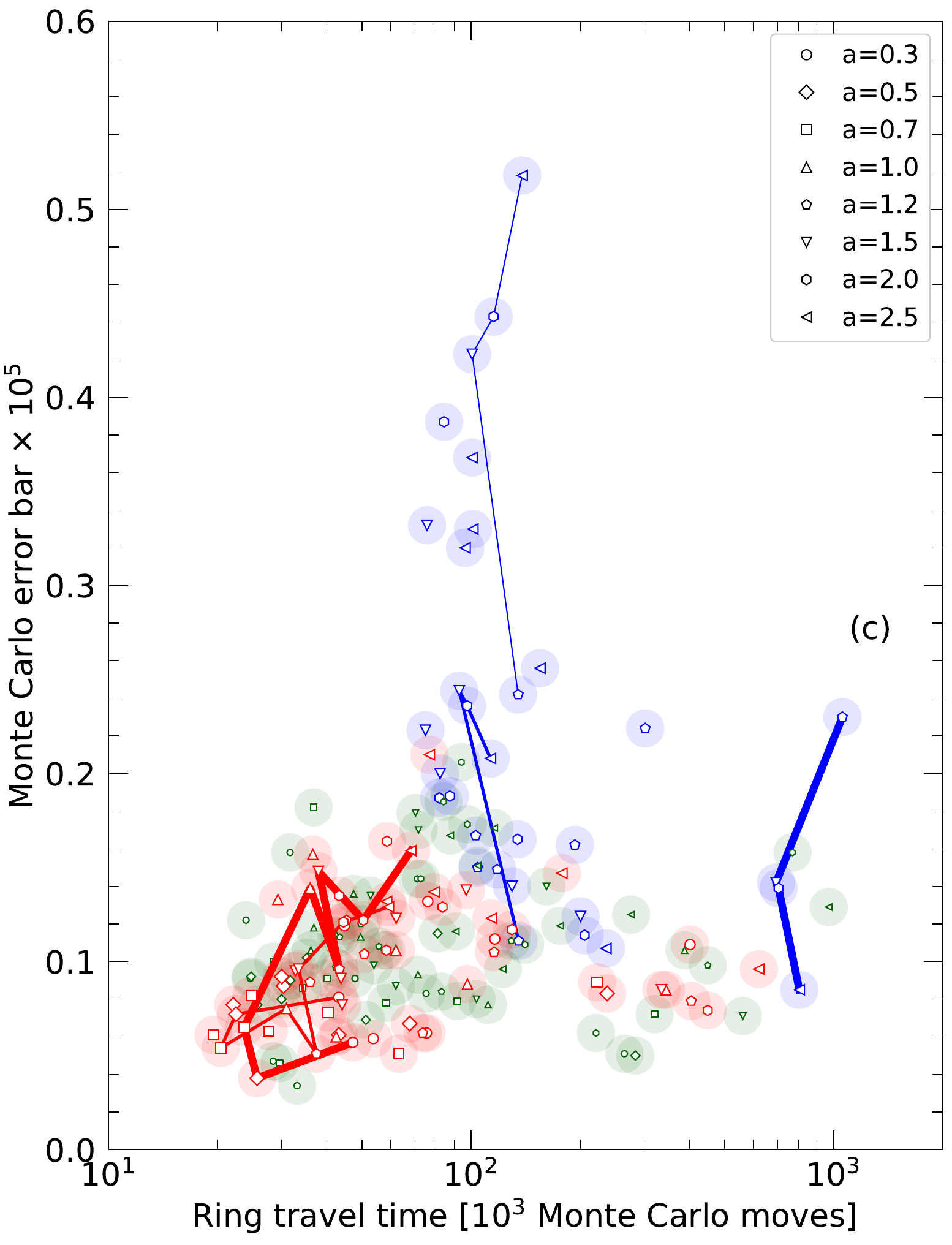}{0.5\textwidth}{}
          }
\caption{\label{fig:ringAuto} Autocorrelation function (panel a) and error bar from blocking analysis (panel b) of affine and quadratic MC calculations. In panel (c), we compare the Monte Carlo error bar from the block analysis and the time it take the ensemble to travel around the ring. The number of attempted MC moves was the same in all cases. The symbols correspond to different stretch factors, $a$, given in the legend. Results for various numbers of walkers are shown, $N_W$=7, 9, 11, 15, 19, 31, 51, and 200. The blue symbols show results derive with the affine method for $a$=1.2, 1.5, 2.0, and 2.5. The thin, medium thick, and thick blue lines represent results with $N_W$=7, 19, and 200 walkers, respectively. The remaining symbols show results from our quadratic MC method with uniform (red) and Gaussian (green) sampling of the $t$ space. The medium thick and thick red lines show the best QMC results for $N_W$=11 and 15, respectively. Compared the affine method, our QMC method requires a approximately travel time half as long and leads to error bars half as large if $N_W = 7 \ldots 19$ and $a = 0.3 \ldots 1.5$ are used.
}
\end{figure}

In Figs.~\ref{fig:ringResults} and \ref{fig:ringAuto}, we compare the
performance the affine MC and our quadratic MC methods under different
conditions. As our baseline case, we set
$N=6, m=6, R=1, C=0.01, T=0.01,$ and $a=2.5$. In every block, we
  attempt to make $10^3$ individual moves. In most cases, we
initialize the ensemble of MC walkers near point $\vec{B}$, which
means average block energy will decrease as the ensemble travels
towards the potential minium near point $\vec{A}$ (see
Fig.~\ref{fig:pot}). In Fig.~\ref{fig:ringResults}a, we compare how
long that takes for different values $m$. Increasing $m$ makes the
potential walls steeper, which causes both methods to converge more
slowly. However, in comparison, the QMC method perform significantly
better. For $m=10$, it only takes 54 blocks for it converge within
2$\times$10$^{-4}$ of the final energy while it takes 308 blocks for
the affine MC method to do so. In Fig.~\ref{fig:ringResults}a, we also
show results from simulations that initialized the ensemble of walker
at the low-energy point $\vec{A}$. The convergence rates are similar
to those before but block averages now converge to the final block
energy from below.

In Fig.~\ref{fig:ringResults}b, we compare the performance of both
method for different ring radii, $R$. For a very small value of 0.2,
both methods converge equally fast. With increasing radius, it takes
the affine MC method much longer than our QMC method to do so. When
we lower the temperature from 0.01 to 0.001, we find a similar
behavior in Fig.~\ref{fig:ringResults}c. A lower temperature makes the
potential appear more confining, which delays the convergence of the
affine MC method dramatically. 

In Fig.~\ref{fig:ringResults}d, we compare the convergence for
different spatial dimensions $N$. For $N=3$ and 6, the QMC method
converges faster but for $N=10$, the behavior is fairly similar to
that of the affine MC method, and it takes both methods longer to
converge than for smaller $N$.

In Fig.~\ref{fig:ringResults}e, we test the dependence on the scaling
parameters $a$. For the QMC method, values between $a=1.1$ and 1.5
yield optimal results. For the affine methods, $a \approx 1.3$ is
optimal but even then it converges only half as fast approximately as
our quadratic method.

Finally in Fig.~\ref{fig:ringResults}f, we vary the temperature that
enters the MC calculation via the Boltzmann factor. Since we are
interested in the effects of ring term in Eq.~\ref{eq:pot} , we set
the constant $C$ equal temperature $T$ for this particular analysis.
A change in $C=T$ recalibrates the strength of the ring term in
Eq.~\ref{eq:pot} in relation to the linear term. For small $C=T$
values, the confining effect of the potential increases, which
foremost delays the convergence of the affine method. 

Summarizing one found that for our ring potential, our QMC method
performed significantly better than the affine MC method for most
conditions. In a few cases like large spatial dimension $N$, the
performance was found to be similar.

In Fig.~\ref{fig:ringAuto}, we study autocorrelations of the block
energy for our base case parameters, $N_W$=7 walkers, and the two
stretch parameter $a=1.5$ and 2.5. For this analysis, we removed the
transient part of the calculation (see Fig.~\ref{fig:ringResults})
where the block energies have not yet
converged. Fig.~\ref{fig:ringAuto}a shows that the autocorrelation
functions of affine MC energies decay much more slowly than that of
the QMC energies. Based on the integrals under the plotted curves, we
estimate the autocorrelation time to be 79000 and 123000 MC moves for
the affine method with a=1.5 and 2.5 respectively but only 12000 and
19000 MC moves for the QMC method (with linear sampling of $P(t)$).

We also performed a block analysis for these four
calculations~\citep{AT87,martin_reining_ceperley_2016}. In
Fig.~\ref{fig:ringAuto}b, we plot the error bars that emerged from the
blocking analysis when individual block energies are combined into
longer and longer blocks. All curves show a plateau that indicates
that the blocks were chosen to be sufficiently long for the block
averages to be uncorrelated. The affine MC method yielded an energy of
$(-3.02 \pm 0.12)\times 10^{-4}$ and $(-3.11 \pm 0.17)\times 10^{-4}$
for $a=1.5$ nd 2.5 respectively.  With the QMC method, we obtained
$(-3.007 \pm 0.051)\times 10^{-4}$ and
$(-3.013 \pm 0.063)\times 10^{-4}$ for the two $a$ values
respectively. All averages are compatible with one another. For the
same calculation duration, the QMC method yielded an error bar that is
2.5 smaller. This is in agreement with observation that its
autocorrelation time is approximately six times shorter.

In Fig.~\ref{fig:ringAuto}c, we compare the performance of the affine
method with that of our QMC method using linear and Gaussian $t$
sampling. It is our goal of this quantitative analysis is give some
guidelines how the stretch factor, $a$, and the numbers of walkers,
$N_W$, should be chosen. Goodman and Weare recommended setting $a=1.5$
and did not make a recommendation for $N_W$ besides choosing it to be
large. (E.g. \citet{Miguel2022} employed 512 walkers to sample a 7
dimensional parameter space.) For the ring potential with $T=0.01$,
$N=6$, $m=6$, $R=1$, we consider value of the stretch factor $a$ from
0.3 to 2.5 to explore the perform of all three methods even though we
consider value $a<0.5$ and $a>1.5$ poor choices. (The affine method
requires $a>1$ while the others do not.)

A large number of walkers introduces diversity, which helps to explore
the parameter space. On the other hand, if the number of walkers is
chose to be too be large, one would expect to algorithm to have
difficulties to explore all relevant areas of the parameter space
efficiently. In principle one would expect the number of walkers scale
with the dimensionality of the space.

In our view, an efficient MC method should have two properties. It
should travel effectively from improbable parameter regions to the
relevant ones. Once there, it should yield to small error bars for the
estimated averages. The two axes of Fig.~\ref{fig:ringAuto}c measure
both properties. On the Y axis, we plot the error bar that we obtained
with the blocking analysis for long MC calculations (one per pair of
$a$ and $N_W$ parameters) with 10$^{10}$ moves. We initialized the
ensemble near low-energy point $\vec{A}$ because, for the error bar
calculations, we are are not interest in the time it take the ensemble
around the ring. To determine the travel time, we performed 10$^3$
separate but shorter calculations with 10$^7$ moves starting from
point $\vec{B}$. Every time, we recorded the ring travel time that we define
to be average number of MC moves that are required for the energy in
the ensemble to reach the mid value between the initial potential
energy and final converged value that we quoted above. The ring travel
time and the MC error bar both have statistical uncertainties, which
introduces noise into Fig.~\ref{fig:ringAuto}c. Still a number of
trends emerge clearly.

If a unreasonably number of walkers like $N_W=200$ is chosen for the
affine method, the ring travel time becomes very large and approaches
10$^6$ MC moves. To a lesser degree, this trend is also seen for the
QMC method. On the other hand, the computed error bars do not suffer
from choosing $N_W$ very large. So employing a very large ensemble of
walkers yields comparable but not smaller error bars than employing
more modest numbers of walkers.

If only 7 walkers are used for the affine method, the ring travel
times become very reasonable but resulting MC error bar becomes very
large. The best performance is seen for $N_W = 9 \ldots 19$ and
$a$=1.5, which yields an average ring travel time of $8 \times 10^4$ moves
and an error bar of $2.5 \times 10^{-6}$.

For the same number of MC moves, our QMC method yields error bars half
the size of the affine method and ring travel times that are
approximately half as long. We see no particular advantage of using
the Gaussian $t$ sampling method. The linear $t$ sampling method
yields the very good results over a wide range $a$ values from
$a=0.3 \ldots 1.5$ and $N_W$=7 $\ldots$ 19. The average travel time is
only $4 \times 10^4$ moves and the error bar is $1.1 \times 10^{-6}$.
Based on this analysis, we recommend setting $N_W=2N+1 \ldots 3N+1$ in
general.

\subsection{Comparison between Quadratic Sampling and Walk Moves}

\begin{figure}[ht!]
\plotone{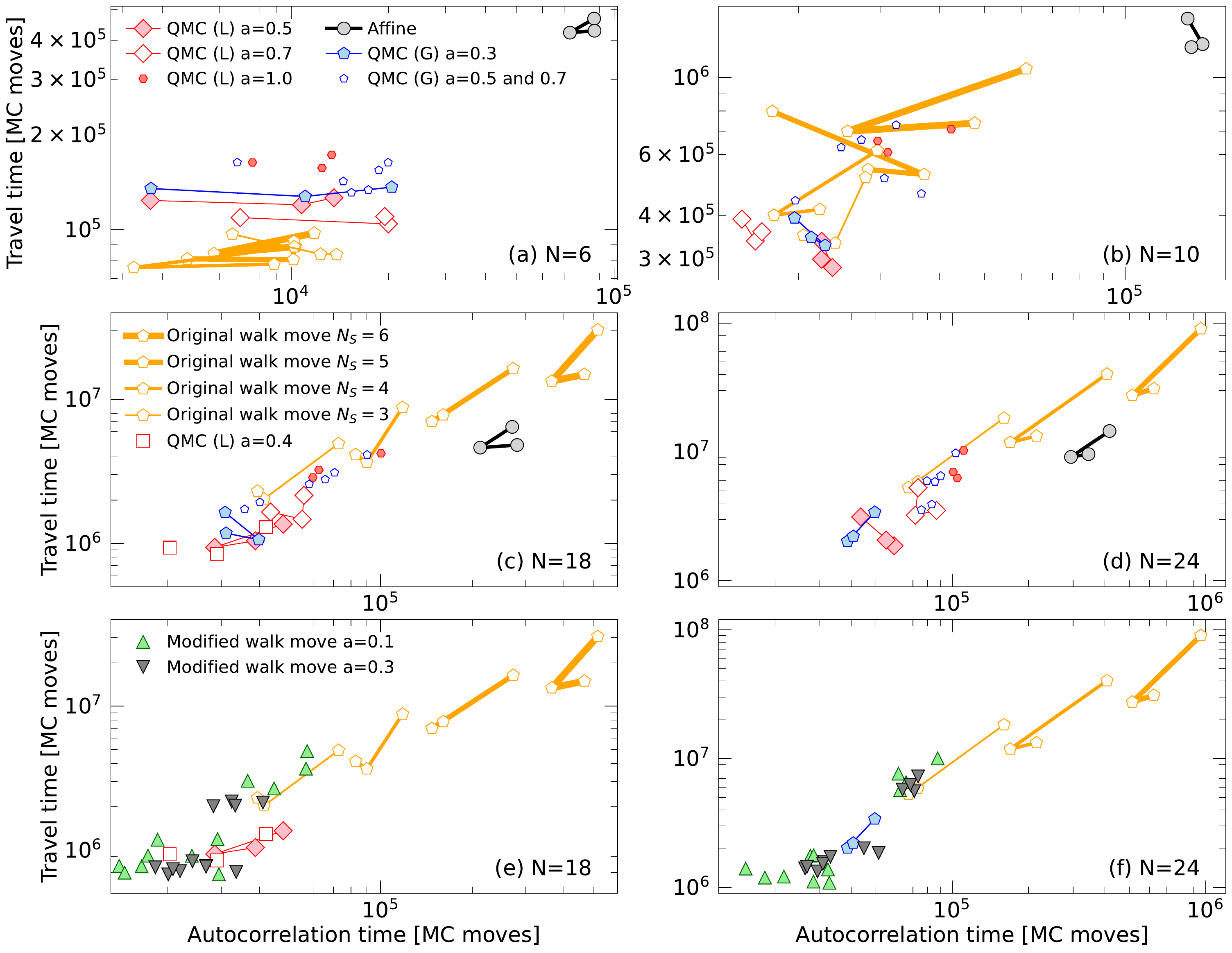}
\caption{Travel and autocorrelation times computed for the ring potential in dimensions $N=6$, 10, 18 and 24. An efficient algorithm makes both as short as possible. Panels (a)-(d) compare results from the affine, quadratic and the original walk method ($a=1$). For a low dimension of $N=6$, the walk moves perform best regardles whether subset size, $N_S=3 \ldots 6$, is chosen but the original walk moves are not competitive for $N>10$. Panels (e) and (f) include results from the modified walk method for $N=18$ and 24 because we found that choosing a scale factor $a \ll 1$ increases the sampling efficiency. Symbols were chosen consistently across all panels. So QMC (L) and (G) label results from the quadratic MC method with linear and Gaussian $t$ sampling respectively. The symbols distinguish results that were obtained with different $a$ parameters.\label{fig:ModifiedWalk}}
\end{figure}

In Fig.~\ref{fig:ModifiedWalk}, we compare the travel and
autocorrelation times from the walk method for the ring potential in
$N=6$, 10, 18 and 24 dimensions with results obtained affine and
quadratic methods using linear and Gaussian $t$ sampling. For every
dimension $N$, we performed independent calculations for
$N_W=N+2, 3N/2,$ and $2N$. For the affine method, we fix $a=1.5$ but
for the quadratic MC method, we considered
$a=\{0.1,0.2,0.3,0.4,0.5,0.7,1.0\}$ for the linear $t$ sampling and
$a=\{0.3,0.5,0.7\}$ for the Gaussian $t$ sampling. For original walk
method, we chose $N_S=3,4,5,$ and 6 for the size of the subset of
guiding walkers. We noticed that choosing $N_S$ larger made such
calculations very inefficient since it led to drastic increases in the
travel and autocorrelation times for $N \ge 10$ as panels (b)-(d) of
Fig.~\ref{fig:ModifiedWalk} illustrate. For lower dimension of $N=6$,
however, the results of the original walk methods are very good. Panel (a)
shows that the travel time can be up to 25\% shorter than that of the
quadratic sampling method.

Already for $N=10$ dimensions results from the original walk
method fall behind those of the quadratic sampling method. For $N=18$
and 24, this trend continues and for a subset size of $N_S=6$, the original walk method
yields longer travel and autocorrelation times than even the affine
method, regardless what ensemble size, $N_W$, is employed. This
increase in travel and autocorrelation times led us to introduce the
scaling factor $a$ into Eq.~\ref{eq:W}. Choosing small values of
$a=0.1$ or 0.3 enabled us to obtained travel and autocorrelation times
with the walk method that are at par or shorter than those of the
quadratic sampling method as panels (e) and (f) of
Fig.~\ref{fig:ModifiedWalk} illustrate. In the next section, we will
analyze how valuable our scaling factor $a$ can be for the sampling of the
Rosenbrock density.

\subsection{Sampling the Rosenbrock Density}

\begin{figure}[ht!]
\gridline{\fig{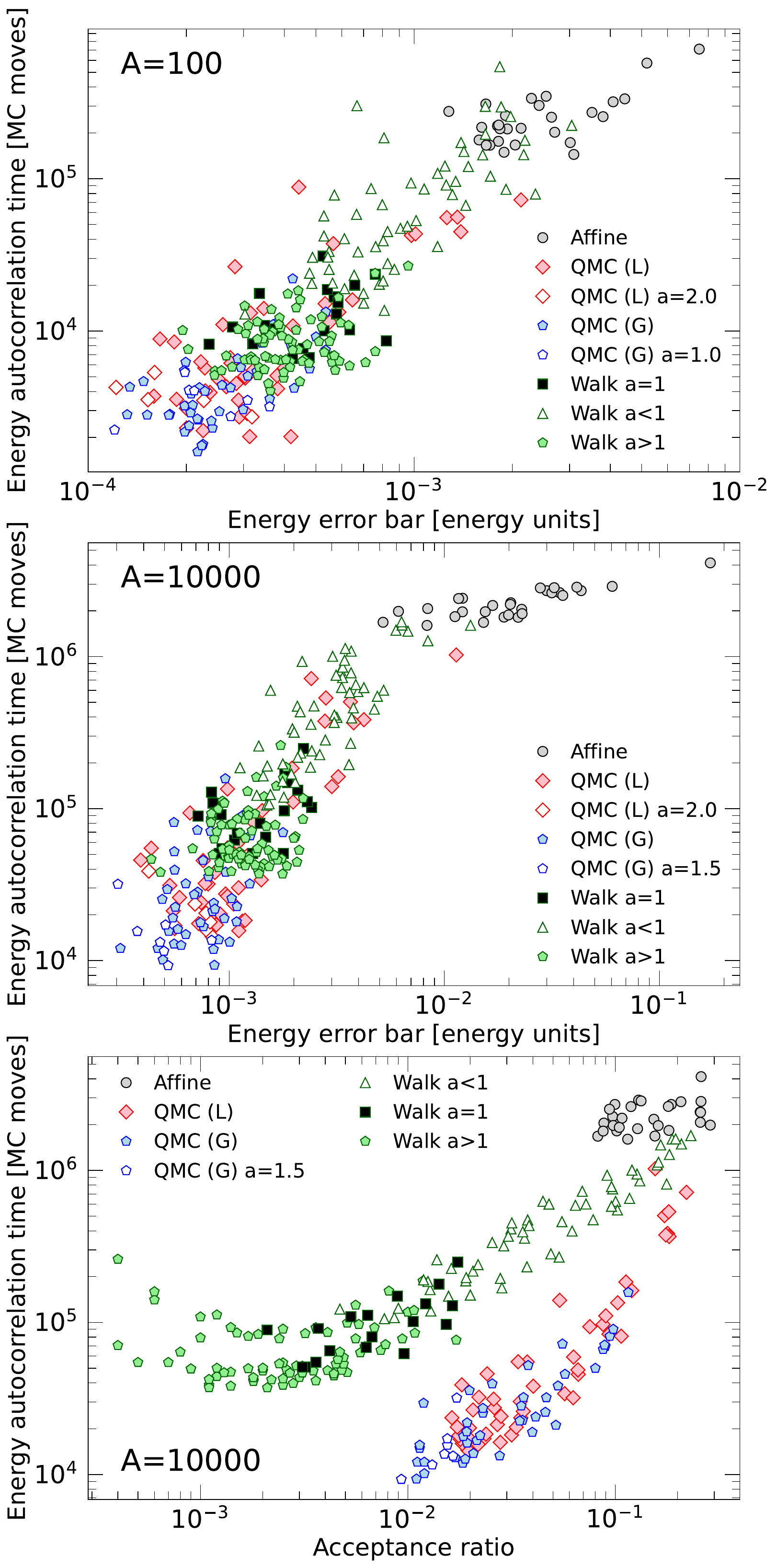}{0.5\textwidth}{}}
\caption{Energy autocorrelation time, energy error bar and acceptance ratios derived for the Rosenbrock density being sampled for $A=100$ and 10000 with the affine method, the quadratic MC method with linear (QMC L) and Gaussian (QMC G) $t$ sampling as well as with the modified walk method. The symbols distinguish results that were obtained with different $a$ parameters.\label{fig:rosen}}
\end{figure}

Following \citet{Goodman}, we also applied our methods to sampling the 2d Rosenbrock density,
\begin{equation}
\pi(x_1,x_2) \propto \exp \left\{ - \frac{A \left( x_2-x_1^2 \right) + (1-x_1)^2}{B} \right\} \quad,
\label{Rosen}
\end{equation}
which carves a narrow curved channel into the $(x_1,x_2)$ landscape.
$B$ effectively plays the role of temperature. First we set $A=100$
and $B=5$ to be consistent with \citet{Goodman} but then we also
increase $A$ to 10000, while leaving $B$ unchanged, which makes the
channel even narrower and makes sampling it yet more challenging.

For both $A$ values, we performed a series of independent MC
calculations with 10$^7$ blocks, each consisting of 10$^3$ individual
moves. We compared the performance of ensembles with
$N_W=\{3,4,5,6,8,10,20\}$ walkers for the following four methods: For
the affine method, we compared the $a$ values $\{1.2,1.5,2.0,2.5\}$,
for the quadratic MC with linear and Gaussian $t$ sampling we
considered $a=\{0.3,0.5,0.7,1.0,1.2,1.5,2.0\}$ respectively. For the
modified walk moves, we studied the combined ranges of
$a=\{0.1,0.3,0.5,1.0,1.2,1.5,2.0,3.0\}$ and $N_S=\{4,5,6,10\}$ under the condition
$N_S<N_W$.

The results are summarized in Fig.~\ref{fig:rosen} where we plot the
autocorrelation time, $\tau$, and the error bar, $\sigma$, that we
computed with the blocking method. Both were derived from average
energy that we computed for everyone of the last 80\% of the 10$^7$
blocks. We define an energy for the Rosenbrock density,
$E(x_1,x_2) = -\ln \pi(x_1,x_2)$, using the analogy between
Eq.~\ref{Rosen} and the Boltzmann factor with $k_bT=1$. 

Despite conserable noise in Fig.~\ref{fig:rosen}, one can identify the
expected scaling of $\tau \sim \sigma^{2}$ between the auto
correlation time, $\tau$, and the computed error bar, $\sigma$. An
optimal algorithm would make both as small as possible. As expected,
both values increase considerably for all algorithms if one raises $A$
from 100 to 10000 because it narrows the channel of the Rosenbrock
density, which makes sampling it yet more difficult.

We find the affine method yields the largest energy error bars among
all methods regardless of which $a$ value is employed. The performance
of the modified walk method strongly depends on the choice of $a$,
which renders the our modification in Eq.~\ref{eq:W} important. For
the sampling of the Rosenbrock density, we find that $a$ values larger
than 1 perform the best, even though they yield an rather low
acceptance ratio of only 2$\times$10$^{-3}$ as the lowest panel of
Fig.~\ref{fig:rosen} illustrates. However, if $a$ is choosen too
large, the acceptance ratio decreases below 10$^{-3}$ and the
autocorrelation time increases because too few of the large steps get
accepted.

We found that the quadratic MC method samples the Rosenbrock density
most efficiently. For $A=10000$, the shortest autocorrelation time
were approximately four times short than the best results that we
obtained with the modified walk method. In Fig. ~\ref{fig:rosen}, we
highlighted some of the most favorable results that were obtained with
Gaussian $t$ sampling for $a=1.5$ and linear $t$ sampling for
$a=2.0$. The acceptance ratio were again rather low and ranged from
0.01 to 0.03 only. This means for challenging sampling problems like
the Rosenbrock density, one may want invest in determining an optimal
or at least a reasonable choice for the scaling parameter $a$.

\subsection{Predictions for Jupiter's Interior} 

\begin{figure}
\plotone{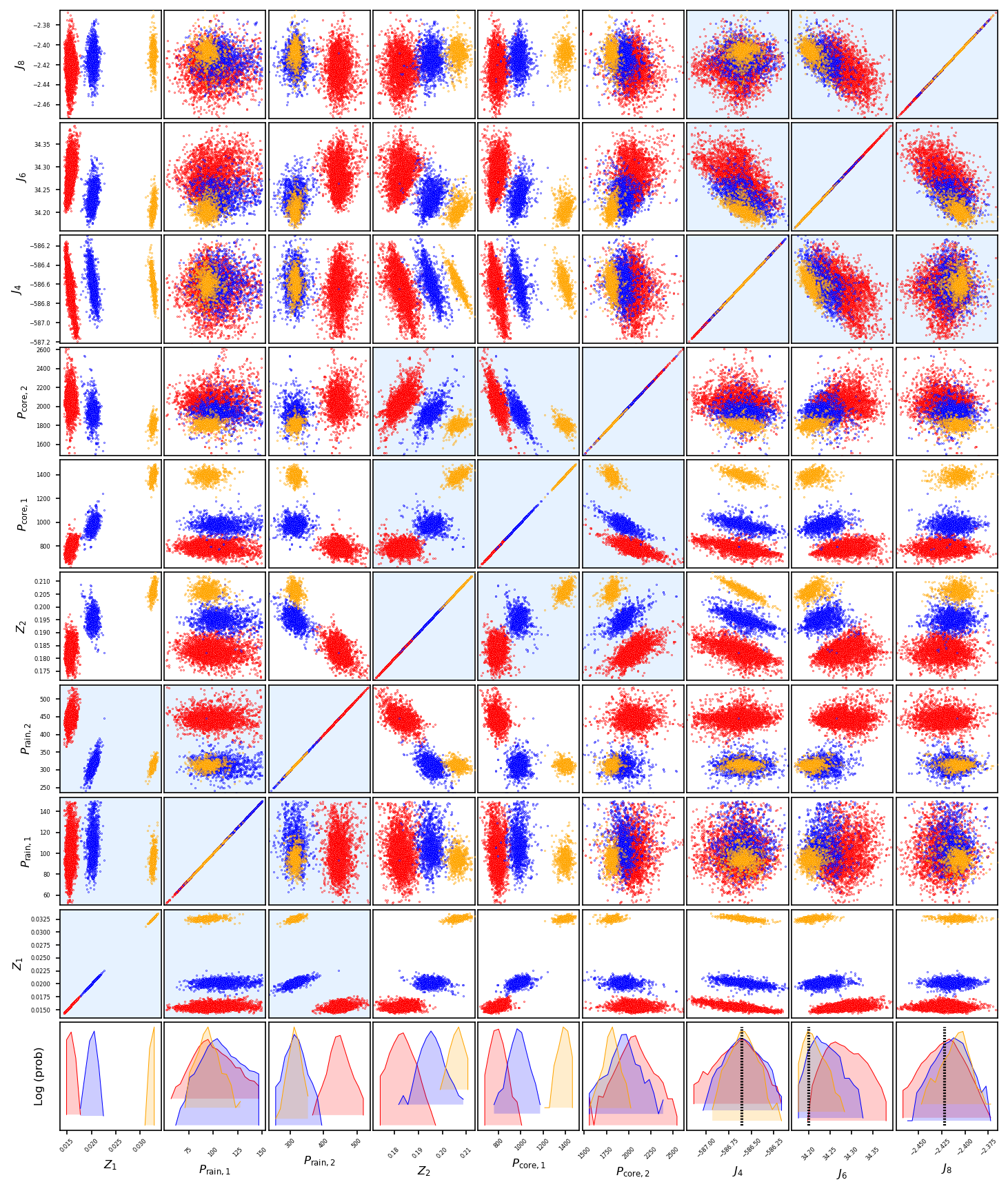}
\caption{Posterior distribution of the three different QMC ensembles: 
1) The red circles represent our reference ensemble of five layer models with a dilute core. 
2) We made Jupiter's interior slightly hotter by increasing temperature at 1 bar from {\em Galileo} measurement of 166.1 to 170 K (blue symbols).
3) We reduced the density of the H-He mixture by 3\% over the pressure interval from 10 to 100 GPa (orange circles).
\label{fig:posterior} $Z_1$ and $Z_2$ are mass fractions of heavy
element. The four pressure values are given in units of GPa. The
values of the gravity harmonics, $J_4 \ldots J_8$, have all been 
multiplied by 10$^6$. The vertical dashed lines indicate the {\em Juno} gravity measurements.\label{posterior1}}
\end{figure}

\begin{figure}
\plotone{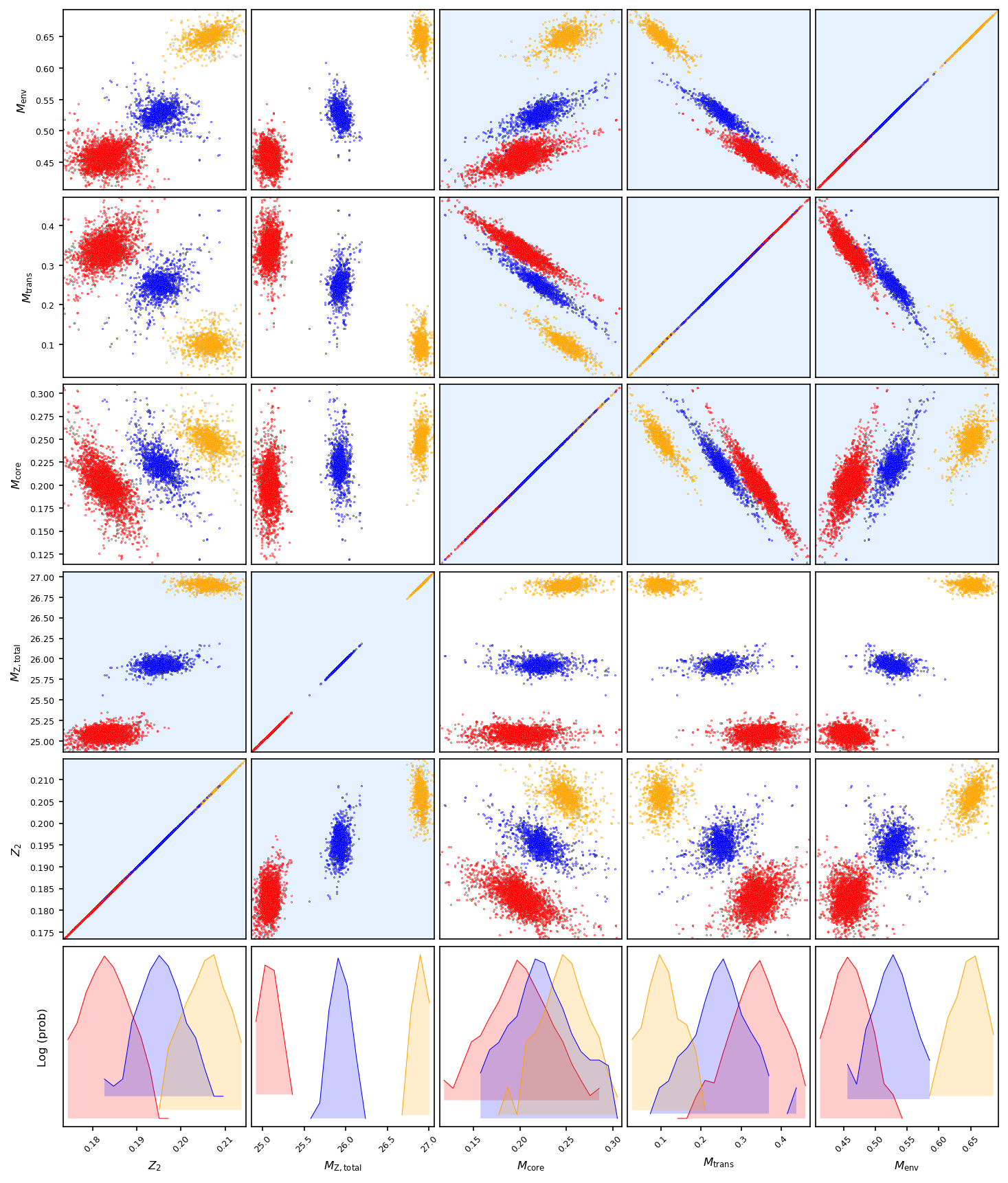}
\caption{Posterior distribution for the three different QMC ensembles
  that we show in Fig.~\ref{posterior1}.  Here we plot correlations
  between heavy elements fraction in the core, $Z_2$, the total amount
  of heavy elements in the planet, $M_{\rm Z,total}$, and the masses
  of the core, the core transition layer and the envelope. (The values
  of three masses were divided by the planet's total mass so that
  they add up to 1.)
  The probability distributions of all variables are shown in the bottom panels on a logarithmic scale.
  \label{posterior2}}
\end{figure}

We applied our QMC algorithm to generate three different ensembles of
interior models under the assuptions in Sec.~\ref{sec:interior}. The
resulting posterior distributions are shown in Figs.~\ref{posterior1}
and \ref{posterior2} while averages and standard deviations of
different parameters are given in Tab.~\ref{tab1}. The three ensembles are:

\begin{enumerate} 
\item This is our reference ensemble of the five-layer models from \citet{DiluteCore}.
\item We increased the
interior entropy by increasing the temperature at 1 bar from the {\em
  Galileo} measurements of 166.1 K to 170 K, which reduces the density
of H-He mixtures in the molecular layer. At the lowest pressures,
where H-He mixture behaves like an ideal gas, this translates into a
density reduction of 2.3\%. At higher pressure, the reduction is
smaller because the systems is more electronically degenerate. 
\item Finally we made a change in our equation of state of H-He
  mixture and reduce the density by 3\% in the region from $P^*=10$ to
  100 GPa but employ a 1 bar temperature of 166.1~K.
\end{enumerate}

\begin{deluxetable*}{cccc}
  \tablenum{1} \tablecaption{Ensemble averages and standard deviations
    of different interior model parameters. Machine-readable data
    files for a representative model of each ensemble are included in
    the supplemental material.
\label{tab1}}
\tablewidth{0pt}
\tablehead{
\colhead{Parameter} & \colhead{Reference} & \colhead{$T_{\rm 1 bar}=170\,$K} & \colhead{3\% density} 
\\
&\colhead{ensemble}&\colhead{ensemble}& \colhead{reduction} }
\decimalcolnumbers
\startdata
$Z_1$ [\%]             & 1.56 $\pm$ 0.05 & 2.03 $\pm$ 0.06 & 3.27 $\pm$ 0.04 \\ 
$P_{\rm rain,1}$ [GPa]  & 98 $\pm$ 16 & 107 $\pm$ 15 & 95 $\pm$ 11\\            
$P_{\rm rain,2}$ [GPa]  & 445 $\pm$ 19 & 314 $\pm$ 19 & 315 $\pm$ 13 \\         
\hline
$Z_2$ [\%]             & 18.3 $\pm$ 0.3 & 19.5 $\pm$ 0.3 & 20.6 $\pm$ 0.3 \\ 
$P_{\rm core,1}$ [GPa]  & 786 $\pm$ 38 & 979 $\pm$ 44 & 1389 $\pm$ 48 \\          
$P_{\rm core,2}$ [GPa]  & 2054 $\pm$ 106 & 1946 $\pm$ 96 & 1811 $\pm$ 63 \\       
\hline
$M_{\rm Z,total} \,[M_E]$ & 25.08 $\pm$ 0.06 & 25.92 $\pm$ 0.06 & 26.90 $\pm$ 0.05 \\
\hline
$M_{\rm core}  \,[M_J]$ & 0.20 $\pm$ 0.02 & 0.22 $\pm$ 0.02 & 0.25 $\pm$ 0.01 \\
$M_{\rm trans} \,[M_J]$ & 0.34 $\pm$ 0.03 & 0.25 $\pm$ 0.03 & 0.10 $\pm$ 0.02 \\
$M_{\rm env} \,[M_J]$   & 0.49 $\pm$ 0.02 & 0.53 $\pm$ 0.02 & 0.65 $\pm$ 0.01 
\enddata
\end{deluxetable*}

Most notably these two density changes increase the mount of heavy
elements, $Z_1$, but they also introduce additional flexibility into
our models and thereby widen the allowed region of other model
parameters, as the larger standard deviations in Tab.~\ref{tab1}
confirm.

One finds that an increase of the 1 bar temperature from 166.1 to 170
K leads to an modest increase in $Z_1$ from $\sim$1.6\% to $\sim$2.0\%
while the 3\% density reduction leads to a much larger increases $Z_1$
to $\sim$3.3\%, effectively doubling the amount. We find increases of
similar magnitude for the heavy elements abundance of the dilute core
region, $Z_2$, from 18.3\% to 19.5\% to 20.6\% when the three
ensembles are compared. Conversely, the ending pressure for the helium
rain layer, $P_{\rm rain,2}$, decreases from $\sim$445 in our
reference ensemble to $\sim$315 GPa in other two ensembles.

Fig.~\ref{posterior1} shows that $Z_1$ is positively correlated with
$P_{\rm rain,2}$ because an increase in $P_{\rm rain,2}$ means helium
is sequestered to deeper layers and the resulting density reduction
over 100--300~GPa pressure interval is compensated by a modest
increase in $Z_1$. In comparison, the correlation between $Z_1$ and
the starting pressure of helium rain layer, $P_{\rm rain,1}$, is
rather weak because typical values for helium rain exponent are
$\alpha \gtrsim 3$ so that the helium concentration does not vary much
near $P_{\rm rain,1}$.  This is also the reason why $P_{\rm rain,1}$
does not strongly correlate with other model parameter.

Fig.~\ref{posterior1} further shows that, within a given ensemble,
$Z_1$ does not correlate strongly with $Z_2$ nor with core pressures,
$P_{\rm core,1}$ and $P_{\rm core,2}$. Still, $Z_1$ is positively
correlated with the magnitudes of $\left| J_4 \right|$, and
$\left| J_6 \right|$ while there is no apparent correlation with
$\left| J_8 \right|$. $Z_2$ and $P_{\rm core,1}$ correlate in the same way
with these three gravity coefficients but strengths of their correlation
are much higher. The extended dilute core is the main feature of our
feature of our five layer models that enables us to fit $J_4$ and
$J_6$ by distributing heavy elements over a wider range of radii than
was possible with compact core assumption. So one expects strong
correlations of $J_4$ and $J_6$ with $Z_2$ and $P_{\rm core,1}$ that
control the heavy element distribution in the core region.

$Z_2$ positively correlates with $P_{\rm core,2}$ because an increase of
$P_{\rm core,2}$ effectively shrinks the size of the dilute core which
is then compensated by an increase in $Z_2$. 
As expected, one finds that $P_{\rm core,1}$ and $P_{\rm core,2}$ are
negatively correlated so the combined mass of heavy elements in the
core and core transition layer is kept approximately constant.

In the bottom row of Fig.~\ref{posterior1} we compare the {\em Juno}
measurements of gravity harmonics $J_4$--$J_8$ with the histogram of
the computed ensembles. $J_4$ is well matched by all three ensembles,
which is a consequence of adopting a dilute core. Matching $J_6$ is
still not straightforward. Models that reduce the density by 3\% are
symmetrically distributed around the measured $J_6$ value. There is
also good overlap with models that adopted a 1~bar temperature of
170~K. Most models with a 1~bar temperature of 166.1~K exhibit a
larger $J_6$ value than was measured. Still as we have shown in
\citet{DiluteCore}, there are models in the 166.1~K ensemble that
match $J_6$ exactly but there are also many others that yield higher
values. In comparison, matching $J_8$ poses no challenge.

In Fig.~\ref{posterior2}, we investigate correlations between the core
mass fraction, $Z_2$, the planet's total budget of heavy elements, and
the masses of the three layers. (Combined they match the planet's
total mass, $M_J$.) When we increase the 1 bar temperature (or lower
the density by 3\% in the 10--100 GPa region), the total amount of
heavy elements increase modestly from 25 to 26 (or 27) Earth
masses. This is a modest increase compared to 8--39 Earth mass range
that \citet{saumon-apj-04} had obtained by considering a plethora of
tabulated EOS models for hydrogen. Our heavy element abundances are a
bit lower than the 28--32 Earth mass range that \citet{Nettelmann2012}
because a compact core and a higher interior temperature profile were
assumed.

When we switch between our three ensembles from 1 to 2 (or to 3), the
mass of the dilute core increases from 0.20 to 0.22 (or to 0.25)
$M_J$. The mass of the core transition layer shrinks drastically from
0.34 to 0.25 (or to 0.10) $M_J$ as $P_{\rm core,1}$ increases and
$P_{\rm core,2}$ decreases. By definition, a rise in $P_{\rm core,1}$
also increases the mass envelope (that include molecular, helium rain,
and metallic hydrogen layers) from 0.49 to 0.53 (or to 0.65)
$M_J$. Switching from ensemble 1 to 2 or 3 leads to a reduction in the
size of the dilute core because one lowers the density in the outer
region of the planet. This is consistent with earlier modeling work
that predicted small or negative heavy elements abundances
\citep{HubbardMilitzer2016} because no dilute core was
considered. 

Fig.~\ref{posterior2} illustartes that, within each ensemble, the mass
of the transition layer negatively correlates with the masses of the
core and that of the envelope. It provides a way to match the total
mass of the planet. More surprising is, however, the masses of the
core and the envelope are positively correlated. This is consistent
with the trend one sees in the first column in
Fig.~\ref{posterior2}. When $Z_2$ increases within a particular
ensemble, the core mass drops, and the envelope mass increases
slightly while the mass of the transition layer remains approximately
unchanged.

\subsection{Equation of state perturbations} \label{sec:EOSperturbations}

Equations of state of materials at high pressure have been studied
with laboratory measurements~\citep{Brygoo2015} and {\em ab initio}
computer simulations~\citep{Mi09,Hu2011,McMahon2012,FPEOS}. At the
same time, it has been a major challenge to the match Jupiter's $J_4$
and $J_6$ with interior models that an rely on a physical equation of
state for H-He mixtures and, for the molecular envelope, yield at
least a protosolar abundance of heavy elements of $Z_{\rm protosolar}=1.53$\% according
to \citet{Lodders2010} who derived the present-day solar abundances in
Tab.~\ref{tab2} by combining spectroscopic measurements of the solar
photosphere with laboratry measurements of CI chondrite
meteorites. Over time, heavy elements diffuse slowly towards a stars
interior because of gravitational forces. \citet{Lodders2010}
represent this process by applying a uniform factor of $10^{0.053}$ to
obtain the protosolar from the solar abundances. Most of the heavy
elements mass comes from just 7 elements that are listed in
Tab.~\ref{tab2}.

\begin{deluxetable*}{cccccc}
\tablenum{2}
\tablecaption{Mass fractions in \% of different heavy elements according to measurements and various models. The second and third column lists the solar and protosolar abundances from \citet{Lodders2010}. The columns four and five show two compositional models for Jupiter. The last column lists one possible interpretation~\citep{HubbardMilitzer2016} of measurement of the {\it Galileo} entry probe~\citep{Wong2004}. \label{tab2}}
\tablewidth{0pt}
\tablehead{
                  & \colhead{Present-day} & \colhead{Inferred}   & \colhead{3-fold proto-} & \colhead{4-fold CO} & \colhead{Galileo} \\
\colhead{Element} & \colhead{solar}       & \colhead{protosolar} & \colhead{solar model  } & \colhead{model for} & \colhead{entry}   \\
                  & \colhead{abundances}  & \colhead{abundances} & \colhead{for Jupiter  } & \colhead{Jupiter  } & \colhead{probe} 
}
\startdata
O      & 0.63 & 0.71 & 2.13 & 1.50 & 0.29 \\
C      & 0.22 & 0.25 & 0.75 & 1.00 & 1.06\\
\hline 
Ne     & 0.17 & 0.19 & 0.02 & 0.02 & 0 \\
\hline 
Fe     & 0.12 & 0.14 & 0.41 & 0.14 & 0 \\
N      & 0.07 & 0.08 & 0.24 & 0.08 & 0.35 \\
Si     & 0.07 & 0.08 & 0.24 & 0.08 & 0 \\
Mg     & 0.06 & 0.07 & 0.20 & 0.07 & 0 \\
Others & 0.07 & 0.08 & 0.24 & 0.08 & 0 \\
\hline
Total  & 1.41 & 1.53 & 4.2 & 2.9 & 1.7 \\ 
\enddata
\tablecomments{Some rounding errors are to be expected.}
\end{deluxetable*}

In Jupiter's atmosphere, the noble gas neon has been measured to be
nine-fold depleted~\citep{Mahaffy2000} compared to the protosolar
abundance. It is assumed that neon partitions strongly into the helium
droplets when hydrogen and helium phase separate at megabar
pressures~\citep{Roulston95,WilsonMilitzer2010}. While the helium
depletion is important for interior models, neon only contribute 11\%
to the solar heavy element budget.

While there is significant uncertainty in the data that have been
obtained for heavy element abundances in Jupiter's atmosphere, one can
make a number of plausible assumptions and then compare them with the
predictions from interior models~\citep{Nettelmann2012}. Here we
compare the predictions from our interior models with three abundance
models in Tab.~\ref{tab2}:

(1) First one can assume all heavy elements are uniformly enriched to
their 3-fold protosolar abundance~\citep{Owen1999} while neon has been
9-fold depleted. This yields $Z^{\rm (3fold)} \approx 4.2\%$. This
assumes the measured enrichment of carbon, nitrogen, and sulfur
applied to all heavy elements even though their respective
condensation temperatures are very different, which may pose a
challenging if one assumes they were delivered along with solid
planetesimals. On the other hand, the near uniform enrichment of the
noble gases suggests that direct capture of nebula gas may have played
a role~\citep{Lodders_2004}. Laboratory condensation
experiments~\citep{NOTESCO2003183} showed the preferred way to
condense noble gases is to trap them in amorphous
ice~\citep{BARNUN2007655} but these measurements also demonstrated the
corresponding trapping rates are nonuniform.

(2) It has also been proposed that oxygen and carbon atoms were delivered in
equal numbers in form of carbon monoxide~\citep{HelledLunine2014}. If one
assumes the measured 4 times protosolar abundance of carbon of reflects this
delivery processes, we obtain $Z^{\rm (CO)} \approx 2.9$\% while we
have included all other elements, except neon, in protosolar
proportions. 

(3) Finally we can take the measurements of the {\em Galileo} entry
probe with its subsolar water abundance at face value,
$Z^{\rm (Gal)} \approx 1.7$\%. While one expects Jupiter's oxygen
abundance to be at least solar, subsolar abundances cannot be ruled
out if Jupiter formed inside the ice line in a region that was starved
of icy planetesimals~\citep{Lodders_2004}. Recently
\citet{Cavalie_2023} predicted a subsolar oxygen abundance for
Jupiter's interior based on thermochemical models for the atmospheres.

In Fig.~\ref{EOS}, we studied how the heavy element abundance in the
atmosphere is affected by an EOS change.  We lowered the H-He density
from~\cite{MH13} by 3\% over a pressure interval from $P^*$ to
$10\times P^*$. The strongest response is found for a $P^*$ range from
0.1 to 3 Mbar, which represent density reductions over broad range of pressure 
(0.1 to 30 Mbar) and includes the transition from molecular to
metallic hydrogen. The resulting models can accommodate more than
double the protosolar abundances in the upper layer. Such an EOS correction can
accomodate the $Z$ abundances of our CO model and get fairly close to matching
the $Z^{\rm (3fold)}$.

Figure~\ref{EOS} also shows that the inferred $Z_1$ value is rather
insensitive to the density change above 3 Mbar where helium rain layer
has ended in most models. This pressure range is also relatively close
to onset of the dilute core, so any change in the H-He EOS may be
compensated by a change in the heavy $Z$ abundance in the core
region. Given this flexibility and the fact that we need the density
to {\em increase} in this pressure interval to match $J_4$ and $J_6$
with a dilute core, explains why $Z_1$ is rather insensitive to a
density correction at such higher pressures.

In Fig.~\ref{EOS}, the vertical lines A and B mark the pressures
where the density of the SC EOS deviates from that of an ideal gas by
respectively 1\% and 10\% because of interaction effects. A density
reduction lead to a modest increase in $Z_1$ only because this region
does not contain a large fraction of the planet's mass.

This leaves the B-to-C region (1-50 kbar). A density reduction by 3\%
there increases $Z_1$ to up 1.75 times the protosolar value. This is
surprising because this region has not yet been studied in sufficient
detail. We are still relying on the SC EOS because the existing
density functional molecular dynamics simulations are not applicable
in this region for two reasons. First, the simulation cells become
very large which makes the expansion of the electronic orbitals in
plane waves very expensive. Second, hydrogen molecules and helium
atoms do not collide very often, which makes it very difficult to
establish a thermodynamic equilibrium within the picosecond time scale
of a typical simulations. Still, Fig.~\ref{EOS} underline this
region should be carefully investigated with theoretical and
experimental method because the predict $Z_1$ is surprisingly
sensitive to the EOS in this pressure region.

\section{Conclusions} \label{sec:conclusions} 

We introduced a novel quadratic Monte Carlo method that performs
significantly better in confined geometries than the earlier affine
(linear) Monte Carlo by~\citet{Goodman}. Both methods rely on an
ensemble of walkers to can adapt to different geometries of the
fitness landscape without manual intervention to guide or improve the
Monte Carlo sampling. 
{There are a number of reasons for why one might want to switch to our
quadratic Monte Carlo method. For a ring potential, we show that our
quadratic Monte Carlo algorithm yields error bars that are half as
large as that of the affine method, which implies that only one
quarter of the computer time is needed to achieve comparable
results. Also our QMC method takes half as long to travel the most
relevant region of parameter space. The discrepancy in efficience
remains present even after the two adjustable sampling parameters, the
number of walkers in the ensemble, $N_W$, and the stretch factor, $a$,
have been optimized for the both methods. We recommend setting $N_W$
between $2N+1$ and $3N+1$ with $N$ being dimensionality of the search
space. We found that choosing $N_W$ much larger increases the time it
takes the ensemble to travel from unfavorable to favorable regions of
the parameter space.

Our QMC method is general and very simple to implement into any
existing MC code. It requires only a few lines of code that we have made available online 
along with examples \citep{QMCOnline}. At the same time,
all applications are different and it remains to be seen whether the
improvements that we report here for the ring potential and Rosenbrock
density carry over to other applications.

We also modified the {\em walk} moves that \citet{Goodman} had
presented as an alternative to the affine invariant moves. We
introduce a new scaling factor, $a$, that enables us to make smaller
(or larger) steps in situations where the covariance of the
instantaneous walker distribution is a not an optimal representation
of local structure of the sampling function. We showed that this
factor improves the sampling efficiency of the Rosenbrock
density. Given the curvature of the its fitness landscape, sampling
this density is particularly challenging for the affine method. The
autocorrelation time of our quadratic Monte Carlo method is two orders
of magnitude shorter.}

We apply our quadratic Monte Carlo method to construct five layer
models of Jupiter's interior that match data from {\em Juno} and {\em
  Galileo} space missions under one set of physical
assumptions. Assuming a dilute core to extends to $\sim$60\% of the
planet's radius enables us to match the gravity field as measured by
the {\em Juno} spacecraft while assuming the helium abundance and 1
bar temperature from the {\em Galileo} entry probe. Constructing
models with a 3-fold enrichment of heavy elements in the planet's
atmosphere remains a challenge unless one invokes an {\em ad hoc}
decrease in the density of hydrogen-helium mixture in pressure range
from 0.1 to 3 megabar where the model predictions are found to be
fairly sensitive. So provide a motivation to revisit the accuracy of
the equations of state of hydrogen and helium with novel experimental
and theoretical methods in the pressure range. On the other hand, an
increase of the 1 bar temperature from 166.1 to 170 K as recently
suggested by \citet{Gupta2022} yields only a modest increase in the
inferred heavy element abundances.

\begin{acknowledgments}
This work was supported by NASA mission {\em Juno} and by the
National Science Foundation's Center for Matter at
Atomic Pressures.
\end{acknowledgments}

\appendix

\section{Proof of Detailed Balance}
\label{proof}

Assuming ergodicity, Monte Carlo simulations are guaranteed to sample
the function, $\pi(\rr)$, in the limit of large step numbers if the
condition of detailed balance is satiesfied (see for example \citet{Ce95}). This
condition is often formulated for transitions between two individual
states $\rr$ and $\rr'$,
\begin{equation}
 \pi(\rr) P(\rr \to \rr') = \pi(\rr') P(\rr \to \rr')
\end{equation}
but here we follow the work by \citet{GreenMira2001} who formulated a generatized condition for detailed balance,
\begin{equation}
 \int \pi(d\rr) P(\rr \to d\rr') = \int \pi(d\rr') P(\rr' \to d\rr)
\end{equation}
where one integrates over states $(\rr$,$\rr') \in A \times B$ that
have been drawn from Borels sets $A$ and $B$, which will be
$\mathcal{R}^N$ for our purposes. The notation $\int \pi(d\rr)$ refers
to the integral,
\begin{equation}
 \int \ldots \pi(d\rr) \equiv \int \ldots p(\rr) d\rr \quad,
\end{equation}
where $p(\rr)$ is the normalized probability density for the
unnormalized distribution function $\pi(\rr)$ (see for example
\citet{Geyer1995}).
\citet{GreenMira2001} showed that the acceptance probability for a
move from $\rr$ to $\rr'$ is given by,
\begin{equation}
A(\rr \to \rr') = \min \left\{ 1 , \frac{\pi(\rr') T'(\LL') } {\pi(\rr')T(\LL) } \left| \frac{\partial(\rr',\LL')}{\partial(\rr,\LL)}\right| \right\}
\quad,
\label{GM}
\end{equation}
where a vector, $\LL$, of $m$ random numbers were drawn from a
density, $T$, to generate the new state $\rr'$ from $\rr$. Similarly,
$\LL'$ refers the $m$ random numbers that are required to generate the
reverse move from $\rr'$ back to $\rr$. In this article, we always use
the same functions for both directions, $T = T'$. The last factor in
Eq.~\ref{GM} refers to the absolute value of Jacobian determinant for
the transformation from $(\rr,\LL)$ to $(\rr',\LL')$ in the product
space of states and random numbers. This term leads to factors
$\lambda^\alpha$ in Eq.~\ref{AAffine} and $\left|w_i\right|^N$ in
Eq.~\ref{AQMC} as we will now show.

For the affine invariant moves, Eq.~\ref{affine} employs a single
random number, $\lambda$, to move from $\rr_i$ to $\rr_i'$. For
reverse move, one needs to set 
\begin{equation}
\lambda'=\frac{1}{\lambda}\quad.
\label{lp}
\end{equation}
For the uniform $\lambda$ sampling, one finds
$T_2(\lambda') / T_2(\lambda) = 1$ but for the sampling function $T_1$
in Eq.~\ref{T1}, one derives the factor
\begin{equation}
T_1(\lambda') / T_1(\lambda) = \lambda \quad. 
\label{T1T1}
\end{equation}

To derive Jacobian determinant, we introduce $r_{ia}$ and $r'_{ib}$
label the $N$ individual elements of state vectors $\rr_i$ and $\rr'_i$. From Eqs.~\ref{affine} and \ref{lp}, one finds,
\begin{equation}
\frac{\partial r'_{ib}   }{\partial r_{ia}  } = \lambda \delta_{ab}\quad, \quad
\frac{\partial \lambda' }{\partial \lambda } = \frac{-1}{\lambda^2}\quad,\quad
\frac{\partial \lambda' }{\partial  r_{ia}  } = 0 \quad{\rm and}\quad
\frac{\partial  r'_{ib}  }{\partial \lambda } = r_{ib} - r_{jb} \quad.\quad
\end{equation}
So the absolute value of the Jacobian determinant becomes
$\lambda^{N-2}$, which explains why one needs to set $\alpha=N-2$ for
the sampling function $T_2$. Because of Eq.~\ref{T1T1}, one needs to
set $\alpha=N-1$ for the sampling function $T_1$.

We now use the same approach to derive the factor $\left|w_i\right|^N$
in Eq.~\ref{AQMC} that specifies the acceptance ratio for a move from
$\rr_i$ to $\rr_i'$ according to Eq.~\ref{mymove}. The forward move
requires two independent random numbers, $\LL=(t_i,t_i')$, while their
roles are interchanged for the reverse move, $\LL'=(t_i',t_i)$, which implies
\begin{equation}
\frac{\partial \lambda'_1 }{\partial \lambda_1 } = 0 \quad, \quad
\frac{\partial \lambda'_2 }{\partial \lambda_2 } = 0 \quad, \quad
\frac{\partial \lambda'_1 }{\partial \lambda_2 } = 1 \quad{\rm and}\quad
\frac{\partial \lambda'_2 }{\partial \lambda_1 } = 1 \quad. \quad
\end{equation}
 The
Jacobian becomes a $(N+2,N+2)$ matrix:
\begin{equation}
J = \frac{\partial(\rr_i',\lambda_1',\lambda_2')}{\partial(\rr_i,\lambda_1,\lambda_2)} = 
\begin{pmatrix}
\frac{\partial r'_{ib}   }{\partial r_{ia}} = w_i \delta_{ab} & \frac{\partial r'_{ib}   }{\partial \lambda_1}  & \frac{\partial r'_{ib}   }{\partial \lambda_2} \\
\frac{\partial \lambda'_{1}}{\partial r_{ia}   } = \frac{\partial \lambda_{2}}{\partial r_{ia}   } &
\frac{\partial \lambda'_1 }{\partial \lambda_1 } = 0 &
\frac{\partial \lambda'_1 }{\partial \lambda_2 } = 1 \\
\frac {\partial \lambda'_{2}}{\partial r_{ia}   } = \frac {\partial \lambda_{1}}{\partial r_{ia}   }&
\frac{\partial \lambda'_2 }{\partial \lambda_1 } = 1 &
\frac{\partial \lambda'_2 }{\partial \lambda_2 } = 0
\end{pmatrix} \quad,
\end{equation}
and its determinant is given by a sum over permutations, $\sigma_k$,
\begin{eqnarray}
|J| = \sum_{\sigma_1 \cdots \sigma_N} 
  \prod_{k=1}^{N} \frac{\partial r'_{i,\sigma_k}   }{\partial \lambda_2} \frac {\partial \lambda_{2}}{\partial r_{i,k}   }
+ \prod_{k=1}^{N} \frac{\partial r'_{i,\sigma_k}   }{\partial \lambda_1} \frac {\partial \lambda_{1}}{\partial r_{i,k}   }
- \prod_{k=1}^{N} w_i \delta_{k,\sigma_k} 
 =  \sum_{\sigma_1 \cdots \sigma_N} \prod_{k=1}^{N} w_i \delta_{k,\sigma_k} = w_i^N
\quad,
\end{eqnarray}
which explains the factor in Eq.~\ref{AQMC}.

\end{document}